\documentclass[fleqn,usenatbib]{mnras}

\usepackage{newtxtext,newtxmath}

\usepackage[T1]{fontenc}
\usepackage{ae,aecompl}

\usepackage{graphicx}	
\usepackage{amsmath}	
\usepackage{amssymb}	

\usepackage[referable]{threeparttablex}
\usepackage{float}
\usepackage[T1]{fontenc}

\newcommand{\lm}{log(M$_*$)}

\newcommand{\rEdd}{$\lambda_\text{Edd}$}	
\newcommand{\zspec}{$z_{\text{spec}}$}
\newcommand{\Lradio}{$\text{L}_{1.4\text{GHz}}$}
\newcommand{\leta}{log($\eta$)}
\newcommand{\ldelta}{log(1+$\delta_{\text{gal}}$)}
\newcommand{\lsfr}{log(SFR)}
\newcommand{\lssfr}{log(sSFR)}
\newcommand{\LAGN}{$\text{L}_{\text{AGN}}$}	
\newcommand{\LEdd}{$\text{L}_{\text{Edd}}$}	
\newcommand{\LIR}{$\text{L}_{\text{IR}}$}	
\newcommand{\fAGN}{f$_{\text{AGN}}$}
\newcommand{\pks}{p$_{\text{ks}}$}

\title[Radio-IR Galaxies and Environment]{The Properties of Radio and Mid-infrared Detected Galaxies and the Effect of Environment on the Co-evolution of AGN and Star Formation at $z \sim 1$}

\author[L Shen]{Lu Shen$^{1,2,3}$,\thanks{Contact e-mail: \href{lushen@ucdavis.edu}{lushen@ucdavis.edu}}
Brian C. Lemaux$^{1}$,
Lori M. Lubin$^{1}$, 
John McKean$^{4}$, Neal A. Miller$^{5}$, 
\newauthor{Debora Pelliccia$^{1,6}$, Christopher D. Fassnacht$^{1}$}, Adam Tomczak$^{1}$, Po-Feng Wu$^{7}$, 
\newauthor{Dale Kocevski$^{8}$, Roy Gal$^{9}$, Denise Hung$^{9}$, Gordon Squires$^{10}$ }
\\
$^{1}$Physics Department, University of California, Davis, One Shields Avenue, Davis, CA 95616, USA \\
$^{2}$CAS Key Laboratory for Research in Galaxies and Cosmology, Department of Astronomy, University of Science and Technology of China, \\Hefei 230026, China  \\
$^{3}$School of Astronomy and Space Sciences, University of Science and Technology of China, Hefei, 230026, China \\
$^{4}$Kapteyn Astronomical Institute, University of Groningen, Groningen, the Netherlands \\
$^{5}$Stevenson University, Department of Mathematics and Physics, 1525 Greenspring Valley Road, Stevenson, MD, 21153, USA \\
$^{6}$Department of Physics and Astronomy, University of California, Riverside, 900 University Ave, Riverside, CA 92521, USA\\
$^{7}$National Astronomical Observatory of Japan, Osawa 2-21-1, Mitaka, Tokyo 181-8588, Japan\\
$^{8}$Colby College, 4000 Mayflower Hill, Waterville, Maine 04901, USA\\
$^{9}$University of Hawai'i, Institute for Astronomy, 2680 Woodlawn Drive, Honolulu, HI 96822, USA\\
$^{10}$Spitzer Science centre, California Institute of Technology, M/S 220-6, 1200 E. California Blvd., Pasadena, CA, 91125, USA
}
\date{Accepted XXX. Received YYY; in original form ZZZ}

\pubyear{}

\begin{document}
\label{firstpage}
\pagerange{\pageref{firstpage}--\pageref{lastpage}}
\maketitle

\begin{abstract}

In this study we investigate 179 radio-IR galaxies drawn from a sample of spectroscopically-confirmed galaxies that are detected in radio and mid-infrared (MIR) in the redshift range of $0.55 \leq z \leq 1.30$ in the Observations of Redshift Evolution in Large Scale Environments (ORELSE) survey. We constrain the Active Galactic Nuclei (AGN) contribution in the total IR luminosity (\fAGN), and estimate the AGN luminosity (\LAGN) and the star formation rate (SFR) using the CIGALE Spectral Energy Distribution (SED) fitting routine. Based on the \fAGN~and radio luminosity, radio-IR galaxies are split into: galaxies that host either high or low \fAGN~AGN (high-/low-\fAGN), and star forming galaxies with little to no AGN activity (SFGs). 
We study the colour, stellar mass, radio luminosity, \LAGN~and SFR properties of the three radio-IR sub-samples, comparing to a spec-IR sample drawn from spectroscopically-confirmed galaxies that are also detected in MIR. 
No significant difference between radio luminosity of these sub-samples was found, which could be due to the combined contribution of radio emission from AGN and star formation.
We find a positive relationship between \LAGN~and specific SFR (sSFR) for both AGN sub-samples, strongly suggesting a co-evolution scenario of AGN and SF in these galaxies. 
A toy model is designed to demonstrate this co-evolution scenario, where we find that, in almost all cases, a rapid quenching timescale is required, which we argue is a signature of AGN quenching. 
The environmental preference for intermediate/infall regions of clusters/groups remains across the co-evolution scenario, which suggests that galaxies might be in an orbital motion around the cluster/group during the scenario.  
\end{abstract}


\section{Introduction}
In the conventional picture of radio galaxy, the dominant source of the radio emission is synchrotron radiation from relativistic electrons accelerated by supernova or powered by AGN, with a subdominant component of free-free radiation from H II regions (e.g. \citealp{Condon1992}). 
Given the origin of the radio emission, two main populations are detected: star-forming galaxies (SFGs) and galaxies with active galactic nuclei (AGN) (see \citealp{Padovani2016} recent review paper). 
Radio AGN (RAGN), traditionally-selected by their radio luminosity (i.e, L$_{1.4GHz} > 10^{24}$), are found to be hosted by red and quiescent galaxies and preferentially reside in the cores of clusters, with their AGN powered by inefficient accretion \citep{Best2005}. Recently, radio AGN have been found hosted by galaxies having on going star formation (SF) (e.g., \citealp{Smolcic2009b, Rees2016, Hardcastle2013, Gurkan2015}). 
This type of RAGN is found to be intrinsically different from traditionally-selected RAGN in that the former is powered by more efficient accretion of cold gas (radiative-mode), while the latter is powered by inefficient accretion of hot gas in the cluster/group core region (jet-mode, e.g. \citealp{Ciotti2010, Best2012}). 

The relationship between AGN power (or black hole accretion rate) and SFR has been investigated for many years with mixed results (see recent reviews \citealp{Alexander2012, Kormendy2013, Heckman2014}). 
In the simplest interpretation this coeval nature means the existing cool gas supply on the host galaxy scale can feed the black hole in the centre of the galaxy at the same time as it allows star formation in the host galaxy. Major mergers, secular and processes (i.e., large galaxy bars and violent disc instabilities) are thought to be responsible for transporting available gas on host galaxy scales to the central regions (e.g. \citealp{Hopkins2010,  Ellison2011}). 
\citet{Gurkan2015} found such a AGN power - SFR relation in RAGN at low redshift ($z < 0.6$). However, a question remains on the role of the AGN in regulating the SF activity, either triggering or suppressing it (e.g., \citealp{Mullaney2012, Rosario2012, Page2012, Harrison2012, Zubovas2013}). Specifically, \citet{Page2012} found that star formation was not observed around black holes in bright X-ray ($L_x > 10^{44}$ ergs s$^{-1}$) galaxies at $1 < z < 3$, which suggests that star formation may be suppressed in the host galaxy of a powerful AGN. 
In addition, \citet{Lemaux2014} studied AGN selected, separately, in X-ray, IR and radio bands over a wider redshift range ($0 < z < 4$) and found that AGN were undergoing starbursts more commonly and vigorously than a matched sample of galaxies without powerful AGN, which they argue that the coeval nature of AGN and SF is actually a strong evidence of the AGN-driven quenching in SFGs. 

To study this relation in radio galaxies, multi-wavelength observations are necessary, in order to probe different structures of AGN and their host galaxies (see \citealp{Padovani2017} and reference therein). 
Observations in the Near-Infrared (NIR) to Mid-Infrared (MIR) range, especially the latter, probe the dust torus close to the AGN (1-10pc, e.g., \citealp{Haas2008, Stern2012, Hickox2018} for recent review). In the Far-Infrared (FIR; 60-700 $\mu$m) range, observed through Herschel \citep{Pilbratt2010}, most of the emission comes from dust that has absorbed a significant fraction of the UV and optical emission from young stars in the host galaxies \citep{Chary2001, Mullaney2012}. 
Therefore, by adding MIR data, it is possible to separate traditionally-selected RAGN and RAGN hosted by SFGs, since it has been found that former have weak MIR emission or optical obscuration from dust \citep{Whysong2004, Ogle2006}. In addition, including FIR data can quantify SFR in these host galaxies without being biased by AGN related contamination. 

The environment in which a galaxy resides is also known to trigger and/or quench star formation, as shown by the relationships between galaxy colour, morphology, stellar mass, and star formation rate (SFR) with various measures of environment (e.g., \citealp{Dressler1980, Peng2010, Grutzbauch2011, Peng2012, Tomczak2019, Lemaux2018b}). In simulation studies of cluster environments, it has been found that galaxies follow a ``delayed - then - rapid'' quenching scenario in which a galaxy spends a considerable amount of time in a group/cluster environment unquenched, followed by a rapid truncation of its star-formation (e.g., \citealp{Wetzel2013, Muzzin2014}). This scenario is consistent with observational studies in the Observations of Redshift Evolution in Large Scale Environments Survey (ORELSE: \citealp{Lubin2009}) at $z \sim 1$ using spectroscopically-confirmed galaxies in a wide dynamical range covering from cores to outskirts of clusters/groups to the field \citep{Lemaux2017, Lemaux2018b, Tomczak2019}, and other studies at similar redshift (e.g., \citealp{Poggianti2009, Balogh2016}). AGN, on the other hand, are known to provide a rapid quenching via radiative winds and large-scale outflows (e.g., \citealp{Gofford2015, Hopkins2016}), which are more powerful in massive stellar mass hosts (e.g., \citealp{Kauffmann2003, Bongiorno2016, Lanzuisi2017, Kaviraj2007}). 

In \citet{Shen2017}, properties of radio galaxies and their environmental preferences in large scale structures at $z \sim 1$ were studied. Radio galaxies were separated into AGN, Hybrid, and SFGs populations. We found that the Hybrid hosts are broadly distributed in colour and stellar mass, with younger stellar ages than the AGN, but older than SFGs. The spectral analyses strongly suggest that they have coeval AGN and SF activity with high accretion efficiency. They do not show clear environmental preferences compared to galaxies of similar colour and stellar mass. 
In this work, we continue our study of the properties of radio galaxies with ongoing SF and AGN activities. We continue to make use of the ORELSE survey to explore radio galaxies at z $\sim$ 1. ORELSE is an extensive photometric and spectroscopic survey of 16 of the most massive large scale structures (LSSs) known at 0.7 $\leq$ z $\leq$ 1.26. With hundreds of high-quality spectroscopic redshifts per field, these data make it possible to accurately map out the three-dimensional density field of each LSS, which reveal a full range of environmental densities at these redshifts (see \citealp{Lemaux2017, Shen2017, Rumbaugh2017, Hung2019}). By combining the ORELSE data with additional mid-infrared imaging from Spitzer/Multiband Imaging Photometer for Spitzer (MIPS: \citealp{Rieke2004}), we can carefully explore the coeval nature of AGN and SF activity in radio galaxies and the role of environment. We use eight fields that have both Very Large Array (VLA) 1.4GHz imaging and Spitzer/MIPS to locate radio-IR galaxies by matching them to a large data set of spectroscopically-confirmed galaxies that are detected in Spitzer/MIPS imaging. 

This chapter is laid out as follows. In Section \ref{sec:Data}, we discuss the data reduction, sample selection and spectral energy distribution (SED) \- fitting, which estimates the AGN contribution to the total IR luminosity (\fAGN), AGN IR luminosity (\LAGN), star formation rate (SFR), and the environment measurements. 
In Section \ref{sec:properties}, we compare the properties of radio-IR galaxies in colour, stellar mass, radio luminosity, \LAGN~and SFR, and environmental preferences. Our main results regarding the \LAGN - SFR relationship are shown in Section \ref{sec:SFR_AGN}. To explain these observations using a physical picture, we devise a toy model and discuss the interpretation of the co-evolution of AGN and SF and the role of environments in Section \ref{sec:discussion}. We conclude with a summary in Section \ref{sec:conclusion}. Throughout this paper all magnitudes, including those in the IR, are presented in the AB system \citep{Oke1983, Fukugita1996}. We adopt a concordance $\Lambda$CDM cosmology with $H_0 = 70~\text{km}~\text{s}^{-1} \text{Mpc}^{-1}$, $\Omega_{\Lambda} = 0.73$, and $\Omega_{M} = 0.27$, and a Chabrier initial mass function (IMF;~\citealp{Chabrier2003}).

\section{DATA AND METHODS}
\label{sec:Data}

In this section, we introduce the observational data and describe the reduction of the optical/near-infrared/mid-infrared photometric, spectroscopic, radio and far-infrared data, the Spectral Energy Distribution (SED) fitting, the method adopted to estimate local and global environment measurements, and the control sample selection method. 
In this paper, we use eight fields (Cl1350, XLSS005, RXJ1053, SG0023, SC1604, RXJ0910, RXJ1716, RXJ1821) from the ORELSE survey, which have fully reduced radio catalogues and are covered by Spitzer/MIPS imaging, along with accompanying photometric and spectroscopic catalogues. We note that we do not include Cl1429 in this study due to the poor radio and Spitzer/MIPS imaging quality for this field.   
Four fields (XLSS005, RXJ1053, SG0023, RXJ0910) have Herschel/SPIRE coverage. 
These observations span 7 $\sim$15 Mpc in the plane of the sky and encompass 32 spectroscopically-confirmed clusters/groups and 97 new overdensity candidates, spanning a total (dynamical) mass range of $10^{12.8} M_\odot$ to $10^{15.1} M_\odot$. See \citet{Rumbaugh2018} and \citet{Hung2019} for more details on clusters/groups and new overdensity candidates in the eight fields.
We summarize the available fields in Table~\ref{tab:fields}, including their central position, redshift, number of clusters/groups and new overdensity candidates, as well as the number of spectroscopically-confirmed and radio galaxies in each field. 

\begin{table*}
	\caption{Properties of ORELSE fields and number of spec-IR/radio-IR sample in each field}
	\label{tab:fields}
	\begin{threeparttable}
	    \begin{tabular}{lcccccccc}
		    \hline
		    \hline
	    	Field & R.A.\tnote{1} & Decl.\tnote{1} & $<z_{\text{spec}}>$\tnote{2} & Num. of C/G & Num. of & Spec-IR \tnote{5} & Radio\tnote{6} & Radio-IR\tnote{7} \\
	    	& (J2000) & (J2000) & & old/new\tnote{3} & Spec\tnote{4} & Galaxies & Galaxies & Galaxies\\
		    \hline
		    Cl1350 & 13:50:48 & +60:07:07 & 0.804 & 3/3 & 336 & 198 & 18 & 10 \\
		    RXJ1716 & 17:16:50 & +67:08:30 & 0.813 & 4/5 & 184 & 97 & 12 & 8\\
		    RXJ1821 & 18:21:38 & +68:27:52 & 0.818 & 2/0 & 94 & 46 & 14 & 10 \\ 
		    SG0023 &  00:24:29 & +04:08:22 & 0.845 & 6/5 & 114 & 82 & 8 & 8 \\
		    SC1604 & 16:04:15 & +43:21:37 & 0.898 & 10/20 & 649 & 431 & 109&75\\
			XLSS005 & 02:27:10 & -04:18:05 & 1.056 & 1/44 & 314 & 159 & 53 & 27\\
			SC0910 & 09:10:45 & +54:22:09 & 1.110 & 4/17 & 201 & 138 & 25 & 19\\
			RXJ1053 & 10:53:40 & +57:35:18 & 1.140 & 2/3 & 196 & 121 & 26 & 22\\
			\hline
			Total &&& 0.920 & 32/97 & 1889 & 1158 & 257 & 179 \\
		    \hline
	    \end{tabular}
	    \begin{tablenotes}
		\item[1] Coordinates are the center of radio imaging.  
		\item[2] Mean spectroscopic redshift of the main structure in each field. The value in the ``Total'' row is the median spectroscopic redshift of the final sample.
		\item[3] The former is the number of clusters and groups that are spectroscopically confirmed using the method presented in \citet{Gal2008}; the latter numbers are found by the new VMC technique presented in \citet{Hung2019}. See details in Section \ref{sec:global_intro}. 
        \item[4] Number of secure spectroscopically-confirmed galaxies in the redshift range 0.55 $\leq$ z $\leq$ 1.3, within $18.5 \le i'/z \le 24.5$ and $\mathrm{M_* \geq 10^{10}M_\odot}$ (see Section~\ref{sec:specobs}). 
        \item[5] Number of galaxies in the spec-IR sample, see the selection in Section~\ref{sec:sample}).
		\item[6] Number of radio sources that are matched to the overall spectroscopically-confirmed galaxies.
		\item[7] Number of radio galaxies detected at $\geq$ 1$\sigma$ in Spitzer/MIPS 24$\mu$m and have a good fit in CIGALE with reduced $\chi^2 \leq$ 10. 
	    \end{tablenotes}
   \end{threeparttable}
\end{table*}

\subsection{Optical/NIR/MIR Imaging and Photometry}
\label{sec:photobs}
Comprehensive photometric catalogues are constructed for all fields studied in this paper. We summarize the available optical and near-infrared (NIR) observations and the reduction process here. 
Deep multi-wavelength imaging are drawn from archival as well as targeted observations from a wide range of facilities, including Subaru, Palomar, CFHT, UKIRT, and Spitzer. The photometry is heterogeneous as the exact combination of bandpasses varies from field to field; most commonly include B, V, R, I, Z, J, K, [3.6], and [4.5]. See Table A1 in \citet{Tomczak2019} for the full lists of all available optical/NIR photometry. 
Photometry for ground-based optical/NIR imaging was obtained by running Source Extractor (SExtractor;~\citealp{Bertin1996}) on point spread function (PSF)-matched images convolved to the image with the worst seeing. Magnitudes were extracted in fixed circular apertures to ensure that the measured colours of galaxies are unbiased by different image quality from image to image. 

Photometry for the Spitzer/Infrared Array Camera (IRAC) and Spitzer/Multiband Imaging Photometer for Spitzer (MIPS: \citealp{Rieke2004}) imaging was obtained by running the package T-PHOT~\citep{Merlin2015}, due to the large point spread function of these data that can blend profiles of nearby sources together and contaminate simple aperture flux measurements. In brief, data were downloaded from the Spitzer Heritage Archive and reduced using the MOPEX software package \citep{Makovoz2006}. For each field separately, MOPEX is run on the individual corrected basic calibrated data (cBCD) frames. First, overlapping frames are background-matched using the overap.pl routine which calculates additive offsets among exposures. Next, mosaicing is performed using the mosaic.pl routine which performs interpolation, outlier rejection, and coaddition of frames. This step also produces coverage and uncertainty maps that show the number of frames and uncertainty per pixel respectively, the latter of which is used with T-PHOT for extracting photometry. See more details in \citet{Tomczak2019, Tomczak2017}. 

Spectral Energy Distribution (SED) fitting is performed in a two-stage process. First, we used the Easy and Accurate Redshifts from Yale (\textsc{EAZY};~\citealp{Brammer2008}) code to estimate photometric redshifts ($z_{phot}$) for galaxies that lacked spectroscopic redshifts. Rest-frame colours are also derived in this step using the best-fit $z_{phot}$ ($z_{spec}$ when available) which are used to classify galaxies as star-forming or quiescent. 
In the second step, we used the Fitting and Assessment of Synthetic Templates (\textsc{FAST}; ~\citealp{Kriek2009}) code to estimate stellar masses as well as other properties of the stellar populations of galaxies. In brief, \textsc{FAST} creates a multi-dimensional cube of model fluxes from a provided stellar population synthesis (SPS) library. Each object in the photometric catalogue is fit by every model in this cube by minimizing $\chi^2$ for each model. We adopt the model that gives the lowest minimum $\chi^2$ as the best-fit. For this we make use of the SPS library presented by \citet{Bruzual2003}, assuming a \citet{Chabrier2003} stellar initial mass function, allowing for dust attenuation following the \citet{Calzetti2000} extinction law. 
A ``use'' flag is designed to identify objects that are poorly detected, saturated, have catastrophic SED fits (having $\chi^2_{reduced} >$ 10)), and/or likely to be foreground stars. We reject galaxies that do not have the ``use'' flag. 
See Section 2.3 of \citet{Tomczak2017} for a more thorough description of these procedures and assumptions. 

\subsection{Spectroscopy}
\label{sec:specobs}
Spectroscopic targets were selected based on the optical imaging in the $r'$, $i'$, and $z'$ from LFC imaging following the methods in~\citet{Lubin2009}.  In brief, the spectroscopic targeting scheme employed a series of colour and magnitude cuts that applied to maximize the number of targets with a high likelihood of being on the cluster/group red sequence at the presumed redshift of the LSS in each field (i.e., priority 1 targets). However, the fraction of priority 1 targets which entered into our final sample ranged from 10\% to 45\% across all ORELSE fields, a fraction which tended to vary strongly with the density of spectroscopic sampling per field (see \citealp{Tomczak2017} for more discussion). 
In addition, for certain masks we prioritized X-ray and radio detected objects.  
The optical spectroscopy was primarily taken with the DEep Imaging and Multi-Object Spectrograph (DEIMOS;~\citealp{Faber2003}) on the Keck II 10m telescope, and reduced using a modified version of the Deep Evolutionary Exploratory Probe 2 (DEEP2, \citealt{Davis2003,Newman2013}) pipeline. See \citet{Lemaux2018b} for details on the modifications to the pipeline. 
In addition to the DEIMOS observations, a notable number of spectroscopic targets (39\%) used in this work are drawn from other literature studies (\citealp{Oke1998, Gal2004, Gioia2004, Tanaka2008}) which utilized different telescopes and instruments. Of these additional targets, the majority (2043) come from the VIMOS VLT Deep Survey (VVDS: \citealp{LeFevre2013}) observations of the XLSS005 field. See \citet{Lemaux2014} for spectroscopic observations in this field. 

Spectroscopic redshifts of these targets were extracted and assessed using the methods of~\citet{Newman2013}, while serendipious detections were added following the method described in~\citet{Lemaux2009}. 
Only galaxies with high quality redshifts (Q=3,4; see \citealt{Gal2008, Newman2013} for the meaning of these values) are used in this study. 
Because the estimated stellar mass completeness limits for these fields range between $10^{9} - 10^{10} M_\odot$ at these redshifts (see \citealp{Tomczak2017}), we apply a stellar mass cut at $\mathrm{M_* \geq 10^{10} M_{\odot}}$ to the overall spectroscopically-confirmed galaxy sample. 
This final spectroscopic sample consists of 1889 galaxies with secure spectral redshifts (i.e. Q=3,4), reliable photometry, and within the adopted stellar mass range (\lm $\leq$ 10). The numbers in the samples for each field are listed in Table \ref{tab:fields}. 

\subsection{Radio Observations and Catalogs}
\label{sec:Radioobs}	

All fields were observed using Karl G. Jansky Very Large Array (VLA) at 1.4GHz in its B configuration, where the resulting FWHM resolution of the synthesized beam is about 5'' and the field of view (i.e., the FWHM of the primary beam) is approximate in 31' diameter. Net integration times were chosen to result in final 1$\sigma$ sensitivities of about 10$\mu$Jy per beam. The details of data reduction and catalogs of the SC1604, SG0023 and RXJ1821 are described in \citet{Shen2017, Shen2019}. Here we describe the details of data reduction and catalogs of the rest of four fields (XLSS005, RXJ1053, RXJ0910 and RXJ1716). 

In general, \texttt{CASA} (Version 4.7.2) was used in calibration and imaging of the VLA data. The standard calibration pipeline was first run on the $(u, v)$ data for each specific observation data set, where obvious interference and other aberrational data were removed, and the resulting gain and phase calibrations were applied to the target fields. 
Then automatic flagging of Radio Frequency Interference (RFI) was conducted via the Time-Frequency Crop (TFCrop) and the rflag procedures, both of which detect and remove outliers in the 2D time-frequency plane. The target fields were imaged using the tclean algorithm with $w$-projection, which is a wide-field imaging technique that takes into account the non-coplanarity of the baselines as a function of distance from the phase center. Each $(u, v)$ data set was self-calibrated. The $(u,v)$ data were then concatenated to produce a single $(u, v)$ data file corresponding to the complete set of observations (i.e., combining all observation dates) for each target field. The final rounds of imaging and self-calibration were then performed on these $(u, v)$ data using the same method for each data set. 

The final images were then used to generate source catalogs. The NRAO's Astronomical Image Processing System (AIPS) task SAD created the initial catalogs by examining all possible sources having peak flux density greater than three times the local RMS noise. We then instructed it to reject all structures for which the Gaussian fitted result had a peak below four times the local RMS noise. Because Gaussian fitting works best for unresolved and marginally resolved sources, residual images created by SAD after having subtracted the Gaussian fits from the input images were inspected in order to adjust the catalog. This step added those extended sources poorly fitted by a Gaussian. Peak flux density, integrated flux density, and their associated flux density errors ($\sigma$) were generated by SAD. We use the peak flux density unless the integrated flux is larger by more than 3$\sigma$ than the peak flux for each individual source. The depths of radio imaging are shown in Table \ref{tab:image_depth}. 

To search for optical counterparts to radio sources, we perform a maximum likelihood ratio (LR) technique, following the procedures in Section 3.4 of \citet{Rumbaugh2012}. 
In brief, a likelihood ratio is defined to estimate the excess likelihood that a given optical source is a genuine match to a given radio source relative to a chance alignment. 
We then carried out a Monte Carlo (MC) simulation to estimate the probability that each optical counterpart is the true match using the likelihood ratios. The threshold for matching to a single or double object is the same as that used in~\citet{Rumbaugh2012}, though in practice in this paper only the highest probability matched optical counterpart was considered for each radio source. 
The optical matching is done to the overall photometric catalogues. We use a search radius of $1\arcsec$, aimed at being inclusive, i.e., not to miss any genuine matches due to instrumental/astrometric/astrophysical effects. 
In this paper, we focus on radio objects which have photometric counterparts with secure spectroscopically-confirmed redshifts and within the redshift range $0.55\leq z \leq1.3$. We refer to these galaxies as radio galaxies. The number of radio galaxies is listed in Table~\ref{tab:fields}. 

\subsection{Far-IR data}
\label{sec:FRIobs}
In this section, we describe the method of estimating FIR flux from the available Herschel/SPIRE 250, 350 and 500 $\mu$m imaging. 
Many different algorithms have been developed to solve the problem of de-blending low-resolution imaging using prior information from high-resolution surveys \citep{Roseboom2010, Wang2014, Hurley2017}. In this study we use the package T-PHOT \citep{Merlin2015}, which implement a maximum likelihood estimation to generate flux density estimates in low-resolution images using galaxy positional priors extracted from a higher-resolution image, to estimate flux density in the Herschel/SPIRE 250, 350, \& 500 $\mu$m channels. 
In general, T-PHOT uses source positions and morphologies of priors from a high-resolution image to obtain photometric analysis of the lower resolution image of the same field. This method was used for measuring galaxy photometry in Spitzer/IRAC and Spitzer/MIPS in this paper by providing real cutouts from the detection image as priors (see \citealp{Tomczak2017, Tomczak2019} for more details). However, due to the large PSF of SPIRE 250, 350, \& 500 $\mu$m bands images, 18'', 25'', \& 37'', respectively, which is a factor of $>$10 more than our complementary images in short wavelengths, the morphology of priors is not necessary. We, therefore, use source positions as unresolved, point-like prior to measure the aperture flux of profiles of nearby sources blended together. 

In brief, the Herschel/SPIRE level 2.5 point source maps were downloaded from the ESA Herschel Science Archive. A list of priors for positions of possible FIR sources is created using all photometric sources that are detected in Spitzer/MIPS 24$\mu$m $\leq1\sigma$ and have ``use'' flags which reject galaxies having poor photometry (see Section \ref{sec:photobs} for details on the ``use'' flag). T-PHOT was performed at the positions of the priors for each of the three SPIRE images. 
Due to relatively shallow depth of the MIPS observations, and for completeness for potential sources detected with Herschel that have lower flux at 24$\mu$m than our detection limit, we use a 1$\sigma$ detection threshold in 24$\mu$m.  
We list in Tables \ref{tab:image_depth} the 3$\sigma$/1$\sigma$ point source completeness limits for the Spitzer/MIPS 24$\mu$m images. 

To estimate the depth of 250, 350 and 500 $\mu$m SPIRE images we employ the following procedure. In each image, objects are masked using a segmentation map generated by the Source Extractor software (SExtractor: \citealp{Bertin1996}) with 3 pixels as minimum area (approximately the same size as the beam size of each image) and 1.5 as a significance detection threshold. We measure the fluxes in 1000 randomly placed apertures of diameter d = 4 pixels (roughly $\sim$ 1.5$\times$ the beam size of the images) in empty sky locations. We then fit a Gaussian to the distribution of these fluxes and adopt the standard deviation as an estimate of the 1$\sigma$ detection limit. 
Across all fields listed in Table \ref{tab:image_depth} we estimate 250, 350 and 500 $\mu$m imaging depths between 6 $\sim$ 23 mJy. We note that the 1$\sigma$ detection limits estimated in this way represent the upper limits in these highly confused images. We also note that by varying the main SExtractor parameters (i.e., minimum area $\pm$ 1 pixels and detection threshold $\pm$ 0.5) would not change the 1$\sigma$ detection limits. 

In Appendix \ref{app:tphot}, we show the performance of T-PHOT using simulation of Herschel SPIRE images. We found that the median of the relative flux difference ($f_{measured} - f_{true})/ f_{true}$) is close to zero, while the scatter of this quantity increases for low-flux objects. We are confident of the fluxes measured down to $\sim$3 mJy where the 16th/84th percentiles of the relative flux difference are $\sim$0.1, which is lower than the $1\sigma$ flux density limits in our Herschel/SPIRE 250 $\mu$m images. See details of this test in Appendix \ref{app:tphot}. 

\subsection{Estimating the AGN contribution and SFR from SED fitting}
\label{sec:CIGALE}

We employed the Code Investigating GALaxy Emission (CIGALE, \citealp{Boquien2019}) in order to constrain the AGN contribution in IR luminosity in a self-consistent framework, considering the energy balance between the UV/optical and IR. 
We adopt a delayed exponential star-formation history (sfhdelayed) allowing $\tau$ and age to range in similar parameter spaces as those used in \citet{Tomczak2019, Tomczak2017}. We assume a \citet{Chabrier2003} initial mass function (IMF) and the stellar population synthesis models presented by \citet{Bruzual2003} with solar metallicity. Dust attenuation follows the \citet{Calzetti2000} extinction law allowing colour excess E(B-V)$_*$ to vary. The reprocessed IR emission of dust absorbed from UV/optical stellar emission is modeled assuming the dust templates of \citet{Dale2014}, allowing the slope ($\alpha$) to vary in a wide range from $\alpha$=1$\sim$2.5 as found in normal star forming galaxies \citep{Dale2002}.

For AGN emission, we utilize the models provided in \citet{Fritz2006}, which take into account two components: the emission of the central source and the radiation from the dusty torus in the vicinity heated by the central source. These models are determined through a set of seven parameters: $R_{max}/R_{min}$, the ratio of the maximum to minimum radii of the dust torus; $\tau_{9.7}$, the optical depth at 9.7 $\mu$m; $\beta$ and $\gamma$ describing the dust density distribution ($\propto r^\beta e^{\gamma|cos\theta|} $) with $r$ the radius and $\theta$ the opening angle of the dust torus; $\Phi$, the angle between the AGN axis and the line of sight; and \fAGN, the AGN contribution in IR luminosity. 
In this paper, we decide to fix the values of $R_{max}/R_{min}$, $\beta$, $\gamma$, $\theta$ that parametrize the density distribution of the dust within the torus to typical values found by \citet{Fritz2006}, in order to limit the number of models. 
We measure the FWHM of the most board emission line (i.e., [OII], [OIII] or [NeIII]) in each spectrum of our radio-IR sample (the selection of this sample is in Section \ref{sec:sample}) and adopt a threshold of $>$500 km/s for potential type 1 AGN. We find a very small fraction (6\%) of our radio-IR sample might host type 1 AGN. Therefore, we fix the $\Phi$ parameter (an angle between the AGN axis and the line of sight) to be 0$^\circ$, corresponding to a Type 2 AGN (i.e. obscured AGN). 
We consider a low ($\tau_{9.7}$= 1) and a high ($\tau_{9.7}$= 6) optical depth model. The former exhibits a smooth distribution in the MIR, while the latter corresponds to a hidden AGN with a strong silicate absorption \citep{Buat2015}. We consider a full range of AGN fraction (\fAGN) from 0 to 0.9, parameterized by the contribution of IR luminosity from AGN to the total IR luminosity \citep{Ciesla2015}. 

For radio synchrotron emission from either star formation or AGN, we consider a correlation coefficient between FIR and radio luminosity ($q_{IR}$) taken from the calculated $q_{IR}$ in our sample, which used the radio luminosities and total infrared luminosities estimated by fitting the IR spectral template introduced by \citet{Wuyts2008} to our measured MIPS 24$\mu$m photometry. The median calculated $q_{IR}$ is 2.12 with the scatter spanning from 1.56 to 2.40 as calculated from the 16th/84th percentiles. 
A wide variety of spectral slopes of power-law synchrotron emission ($\alpha_{\text{radio}}$) is adopted, including the average spectral slope observed for galaxies hosting a RAGN ($\alpha$ = 0.7, i.e., \citealp{Condon1992}), and extremely steep slope ($\alpha > 1$) for galaxies hosting obscured AGN or originating from star-forming regions \citep{Ibar2010}. 

Finally, we adopt the ``pdf\_analysis'' analysis method in CIGALE to compute the likelihood ($\chi^2$) for all the possible combinations of parameters and generate the marginalized probability distribution function (PDF) for each parameter and each galaxy (see Section 4.3 in \citealp{Boquien2019}, for full explanation of this method). More details of the parameter settings are shown in Table \ref{tab:CIGALE}. 

We run CIGALE on photometry measured from Spitzer/MIPS 24$\mu$m, Herschel/SPIRE 250, 350, and 500$\mu$m observations if available, radio 1.4GHz, as well as all available optical and NIR data. The median $\chi^2_{reduced}$ of the best-fit SEDs of the radio-IR sample is 2.69. We define galaxies have $\chi^2_{reduced} >10$ as bad fits, which removes 22 (10\%) of the radio-IR targets. After this cut, the median $\chi^2_{reduced}$ is reduced to 2.24 in the final radio-IR sample. 
In this work, we use SFR, AGN contribution to IR luminosity (\fAGN) and AGN luminosity (\LAGN) estimated from CIGALE. To avoid introducing a bias we use the stellar mass estimated from FAST, instead of that estimated from CIGALE, so that we could compare the stellar mass in the radio-IR sample to the overall spectroscopically-confirmed sample. 
Moreover, in practice this choice does not matter, as the difference of stellar mass estimated from FAST and that from CIGALE is negligible. Even when the AGN component is considered, the difference does not change as a function of stellar mass. We show the difference of two estimated stellar masses in the Appendix \ref{app:CIGALE_compare}. In addition, we further discuss the possible uncertainties and bias of physical properties derived by CIGALE in Appendix \ref{app:CIGALE_compare}. 
Due to the large uncertainties of Herschel/SPIRE 250, 350, and 500$\mu$m photometry, we do not find significant differences when comparing CIGALE best-fit quantities with or without these data. Thus, we treat galaxies equally regardless of Herschel/SPIRE coverage. 

\begin{table}
	\caption{Parameter ranges used in the SED fitting with CIGALE.}
	\label{tab:CIGALE}
	\begin{threeparttable}
	    \begin{tabular}{l|c} 
		    \hline
		    \hline
	    	Parameter  & Value \\
		    \hline
		    \multicolumn{2}{c}{sfhdelayed}\\
		    \hline
         	$\tau_{\text{main}}$[Gyr] & 0.3, 0.5, 0.7, 1, 2, 4, 6, 8, 10\\
         	age[Myr] & 50, 100, 200, 300, 400, 500, 750, 1000, \\
         	&3000, 5000, 8000\\
         	\hline
         	\multicolumn{2}{c}{SSP \citep{Bruzual2003}}\\
         	\hline
         	IMF & 1 \\
         	metallicity & 0.02 \\
         	\hline
         	\multicolumn{2}{c}{Dust attenuation \citep{Calzetti2000}} \\
         	\hline
            E(B-V)$_*$ & 0.01, 0.05, 0.1, 0.15, 0.2, 0.25, 0.3 \\
         	 & 0.4, 0.5, 0.6, 0.8, 1.2\\
         	E(B-V)$_\text{factor}$ & 0.44 \\
         	\hline
         	\multicolumn{2}{c}{Dust emission \citep{Dale2014}} \\
         	\hline
         	alpha & 0.25, 0.5, 1.0, 1.5, 2.0, 2.5 \\
         	\hline
         	\multicolumn{2}{c}{AGN emission \citep{Fritz2006}\tnote{1}} \\
         	\hline
         	$R_{max}/R_{min}$ & 60.0 \\
         	$\tau_{9.7\mu m}$ & 1.0, 6.0 \\
         	$\beta$ & -0.5 \\
         	$\gamma$ & 0.0 \\
         	$\theta$ & 100$^\circ$\\
         	$\phi$ & 0.001$^\circ$  \\
         	\fAGN & 0.0, 0.05, 0.1, 0.2, 0.3, \\
         	& 0.4, 0.5, 0.6, 0.7, 0.8, 0.9 \\
         	\hline
         	\multicolumn{2}{c}{Radio synchrotron emission} \\
         	\hline
         	$q_\text{IR}$ & 1.0, 1.5, 2.0, 2.5, 3.0 \\
         	$\alpha_{\text{radio}}$ & 0.5, 0.7, 0.9, 1.1 \\
         	\hline
		    \hline
	    \end{tabular}
	    \begin{tablenotes}
		\item[1] See Section \ref{sec:CIGALE} for details of these parameters.
		\end{tablenotes}
   \end{threeparttable}
\end{table}

\subsection{Sample Selection}
\label{sec:sample}
With the observational data and the catalogs in hand, we select our sample to have galaxies with:

\begin{enumerate}
    \item secure spectroscopic redshifts between 0.55 and 1.30; 
    \item good fits from our comprehensive photometric catalogs and detected at $\geq$ 1$\sigma$ Spitzer/MIPS 24$\mu$m imaging for reliable estimation of star formation rates and AGN contribution. See details in Section \ref{sec:photobs};  
    \item matched to $\geq$ 4$\sigma$ radio detection source using a Monte Carlo maximum likelihood ratio technique (See details in Section \ref{sec:Radioobs});  
    \item M$_*$ $\geq$ 10$^{10}$ M${_\odot}$ due to the completeness of our spectroscopic sample (see more in Section \ref{sec:specobs}); 
    \item good fits from CIGALE ($\chi_{reduced}^2 \leq$ 10, see Section \ref{sec:CIGALE}).
\end{enumerate}

We obtain a total of 179 galaxies that meet all of the above criteria, called the ``\textbf{radio-IR}'' sample hereafter. For comparison, we also construct a spectroscopic sample that meets criteria (1), (2) and (4) as the parent sample, called the ``\textbf{spec-IR}'' sample hereafter. Note that the radio-IR sample is included in the spec-IR sample. The number of galaxies in the spec-IR and the radio-IR samples in each field are listed in Table \ref{tab:fields}. 

In CIGALE, the IR luminosity contributed by the AGN is defined as \LAGN = \fAGN $\times$ \LIR~\citep{Ciesla2015}, where \LAGN~is the AGN luminosity at IR wavelengths and \LIR~is the total IR luminosity. We divided the overall radio-IR sample based on the estimated \fAGN~and radio luminosity into three sub-samples: SFGs, low- and high-\fAGN, where SFGs are selected to be star formation dominated galaxies that are best fitted by \fAGN~$<$ 0.1, i.e., with little to no AGN activity. 
We also include a radio luminosity threshold here since traditionally-selected RAGN are always found to have \Lradio~$> 10^{23.8}$ (\citealp{Shen2017}, and references therein). This way, the SFG sample is considered to contain pure star-forming dominated galaxies. Note that 9 galaxies are excluded by the radio luminosity threshold. 
Low- and high-\fAGN~are selected to be galaxies that host AGN where AGN contribution in IR luminosity is high (\fAGN~$\geq0.5$) and low ($0.1\leq$\fAGN~$<0.5$). 
The criteria and the number of galaxies are shown in Table \ref{tab:subsamples}. 

\begin{table}
	\caption{Criteria of sub-samples and the number of galaxies.}
	\label{tab:subsamples}
	\begin{threeparttable}
	\begin{tabular}{ccc} 
        \hline
        sub-samples & Criteria & num. of gals \\
        \hline
        \hline
        SFG & \fAGN $<$ 0.1 \& \Lradio $ < 10^{23.8}$ & 70\\ 
        low-\fAGN & 0.1 $\leq$ \fAGN $<$ 0.5 & 54\\
        high-\fAGN & \fAGN $\geq$ 0.5 & 46\\
        \hline
        \end{tabular}
    \end{threeparttable}
\end{table}

\begin{figure}
    \includegraphics[width=\columnwidth]{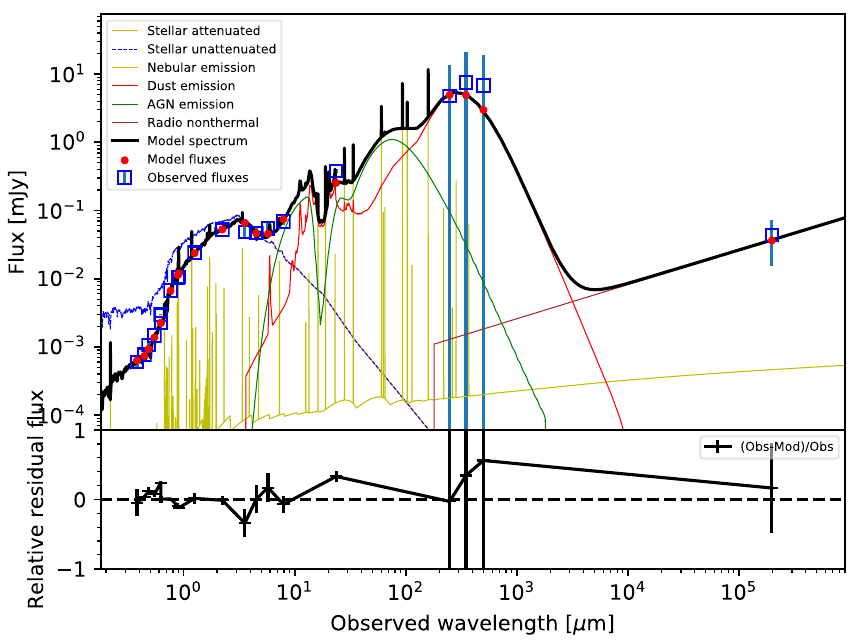}
    \caption{An illustration of the data quality and CIGALE SED modeling. \textit{Top:} The observed photometric fluxes from one radio galaxy (id = 40319) at $z$ = 0.80 in RXJ1053 with errors are shown in blue symbols. The reduced $\chi^2$ is 1.71. The best-fit CIGALE model is shown in black. Red dots indicate CIGALE-derived photometry in the modeled passbands. The best-fit CIGALE model is the sum of contributions from an AGN (green dashed line), dust-attenuated stellar emission (orange; the intrinsic stellar emission is indicated in blue), nebular emission (yellow), and dust emission (red). The \textit{bottom} panel shows the fractional discrepancies between the model and photometry. Note that the best-fit \fAGN~is 0.4. }
    \label{fig:cigale}
\end{figure}

\subsection{Environmental Measurements}
\label{sec:env}

\citet{Shen2017} have shown that different types of radio galaxies have different environmental preferences that suggest different scenarios. The AGN are preferentially located in the cores of clusters/groups, suggesting Bondi accretion to be the possible triggering process, while the SFG population exhibits a strong preference for intermediate regions of the clusters/groups, suggesting they could be driven by galaxy-galaxy interactions and merging. Hybrids do not show clear environmental preferences compared to galaxies of similar colour and stellar mass. 
To further investigate the effect of environment in radio-IR galaxies, we continue to adopt two environment measurements following \citet{Shen2017}: local environment which probes the current density field to which a galaxy is subject, and global environment which probes the time-averaged galaxy density to which a galaxy has been exposed. 

\subsubsection{Local Environmental Density}
\label{sec:local_intro}
We adopt a local environment measurement using a Voronoi Monte-Carlo (VMC) algorithm which is described in full detail in \citet{Lemaux2017} and \citet{Tomczak2017}.  
In brief, 78 thin redshift slices ($\Delta v = \pm 1500~\mathrm{km/s}$) are constructed across the broad redshift range of 0.55 $\leq$ z $\leq$1.3, with adjacent slices overlapping by the half depth of the slice. Spectroscopically-confirmed galaxies are then placed into these thin redshift slices. For each slice, photometric galaxies without a high quality $\mathrm{z_{spec}}$ (with a good use flag) have their original $\mathrm{z_{phot}}$ perturbed by an asymmetric Gaussian with a mean and dispersion set to the original $\mathrm{z_{phot}}$ and $\pm1\sigma$ uncertainty, respectively, and the objects whose new $\mathrm{z_{phot}}$ fall in the redshift slice are determined. 
Then a Voronoi tessellation is performed on the slice and is sampled by a two dimensional grid of 75 $\times$ 75 proper kpc pixels. The local density is defined as the inverse of the cell area multiplied by the square of the angular diameter distance. 
The final density map of the slice is computed by median combining the density values for each pixel from 100 VMC realizations. 
The local overdensity value for each pixel point (i, j) is then computed as $\mathrm{log(1 + \delta_{gal}) = log(1 + (\Sigma_{VMC} - \tilde{\Sigma}_{VMC})/\tilde{\Sigma}_{VMC})}$, where $\mathrm{\tilde{\Sigma}_{VMC}}$ is the median $\mathrm{\Sigma_{VMC}}$ for all grid points over which the map is defined. Local overdensity rather than local density is adopted to mitigate issues of sample selection and differential bias on redshift. 

\subsubsection{Global Environmental Density}
\label{sec:global_intro}
To quantify the global environment, we adopt $\mathrm{R_{proj}/R_{200}}$ versus $|\Delta v|/\sigma_v$~\citep{Carlberg1997, Balogh1999, Biviano2002, Haines2012, Noble2013, Noble2016}, defined by $\eta$ in \citet{Shen2017} as:  
\begin{equation}  
\label{eq:eta}
\mathrm{\eta =(R_{proj}/R_{200})}\times(|\Delta v|/\sigma_v)
\end{equation} 
where $\mathrm{R_{proj}}$ is the distance of each galaxy to each group/cluster center, $\mathrm{R_{200}}$ is the radius at which the cluster density is 200 times the critical density, $\Delta v$ is the difference between each galaxy velocity and the systemic velocity of the cluster, and $\sigma_v$ is the line-of-sight (LOS) velocity dispersion of the cluster member galaxies (see \citealt{Lemaux2012} for the computation of $\Delta v$ and $\sigma_v$). 
The final cluster/group catalog is derived from the previous spectroscopically-confirmed clusters and groups in the ORELSE survey \citep{Ascaso2014} combined with new overdensity candidates \citep{Hung2019}. In the former case, clusters and groups are spectroscopically confirmed using the method presented in \citet{Gal2008}. The cluster centers are obtained from the i$^\prime$-luminosity-weighted center of the members galaxies as described in \citet{Ascaso2014}. 
In the latter case, new overdensity candidates are found using VMC maps (see Section \ref{sec:local_intro}), down to total overdensity masses of log(M$_{\text{tot}}$/M$_\odot$) $>$ 13.5. The $\sigma_v$ of these new structures are velocity dispersions calculated from log(M$_{\text{tot}})$ according to equation 1 and 2 in \citet{Lemaux2012}. 
The value of $\eta$ for each galaxy is measured with respect to the closest cluster/group. To determine it, we first find all the clusters/groups that are within $\pm$6000 km s$^{-1}$ in velocity space of a given galaxy. We then compute $\mathrm{R_{proj}/R_{200}}$ from the galaxy to the clusters/groups and select the one for which $\mathrm{R_{proj}/R_{200}}$ is the smallest as the parent cluster/group. If for a given galaxy no clusters/groups within $\pm$6000 km s$^{-1}$ are found, $\eta$ is computed with respect to all of those clusters/groups in the field and the one with the smallest value is associated with that galaxy. See \citet{Pelliccia2019} for more detail on this calculation.   
In this paper, we separate galaxies to be in the clusters/groups core ($\eta \leq 0.1$), intermediate region ($0.1 < \eta \leq 0.4$), infall region ($0.4 < \eta \leq 2$), and field region ($\eta > 2$). 

\section{Galaxy Properties of radio-IR sub-samples}
\label{sec:properties}

Radio galaxies detected in the MIR are divided into SFGs, low-, and high-\fAGN~sub-samples as described in Section \ref{sec:sample}. 
In this section, we present an analysis of different properties of each radio-IR galaxy sub-sample. We first show the distributions of the radio host galaxies in terms of their optical colour, stellar mass, redshift, and environments, as compared to the overall spec-IR sample. We then present internal comparisons of radio luminosity, star formation rate (SFR), AGN luminosity (\LAGN) between each radio-IR sub-sample, which imply a correlation between SFR and AGN luminosity. Median values of these properties for each sub-sample, radio-IR and spec-IR are listed in Table \ref{tab:properties}.

\begin{table*}
	\caption{Median properties of galaxies in sub-samples, the overall radio-IR and spec-IR samples}
	\label{tab:properties}
	\begin{threeparttable}
	    \begin{tabular}{c|cccccccc} 
        \hline
        Sample & z & colour offset & \lm & \ldelta & \leta & log(\Lradio) & \lsfr & log(\LAGN) \\
        \hline
        \hline
        high-\fAGN & 0.95 $\pm$ 0.03 & -0.40 $\pm$ 0.04 & 10.86 $\pm$ 0.03 & 0.48 $\pm$ 0.04 & -0.38 $\pm$ 0.08 & 23.35 $\pm$ 0.04 & 0.21 $\pm$ 0.07 & 11.04 $\pm$ 0.05 \\
        low-\fAGN & 0.92 $\pm$ 0.02 & -0.37 $\pm$ 0.04 & 10.71 $\pm$ 0.03 & 0.35 $\pm$ 0.03 & -0.61 $\pm$ 0.09 & 23.43 $\pm$ 0.03 & 0.77 $\pm$ 0.04 & 10.40 $\pm$ 0.05\\
        SFGs & 0.90 $\pm$ 0.01 & -0.67 $\pm$ 0.06 & 10.65 $\pm$ 0.04 & 0.47 $\pm$ 0.03 & -0.45 $\pm$ 0.07 & 23.30 $\pm$ 0.02 & 1.02 $\pm$ 0.05 & - \\
        radio-IR\tnote{1} & 0.92 $\pm$ 0.01 & -0.47 $\pm$ 0.02 & 10.73 $\pm$ 0.02 & 0.45 $\pm$ 0.02 & -0.48 $\pm$ 0.04 & 23.36 $\pm$ 0.02 & 0.77 $\pm$ 0.04 & - \\
        spec-IR & 0.90 $\pm$ 0.003 & -0.42 $\pm$ 0.01 & 10.58 $\pm$ 0.01 & 0.37 $\pm$ 0.01 & -0.43 $\pm$ 0.02& - & - & - \\
        \hline
        \end{tabular}
        \begin{tablenotes}
		\item[1] the overall radio-IR galaxies. Note that radio-IR galaxies with \fAGN $<$ 0.1 and \Lradio $> 10^{23}$ are not included in any of the above three sub-samples.
		\end{tablenotes}
    \end{threeparttable}
\end{table*}

\begin{figure}
\centering
    \includegraphics[width=\columnwidth]{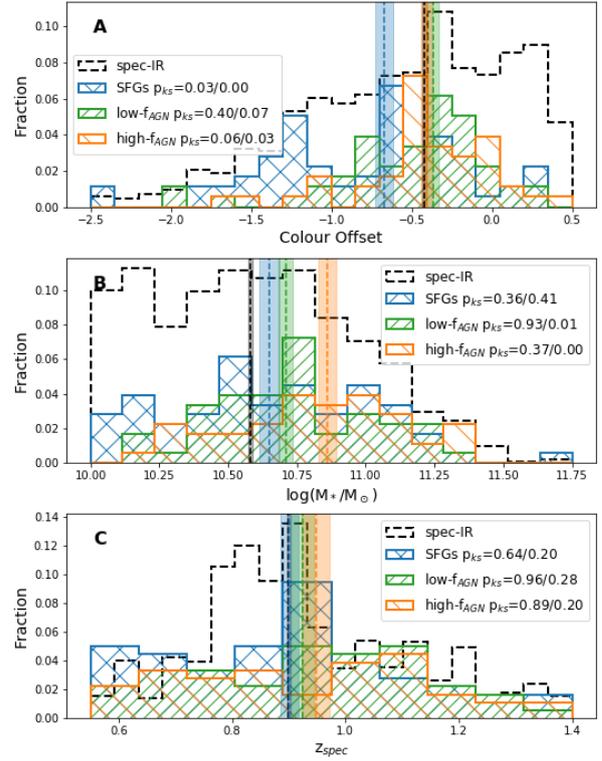}
    \caption{\textit{Panel A:} colour offset histograms of spec-IR (dashed black) and three radio-IR high-\fAGN (orange), low-\fAGN (green) and SFGs (blue) sub-samples, scaled by the total number of Radio-IR sample. A positive colour offset value represents galaxies in the quiescent region, see Section \ref{sec:color}. \textit{Panel B:} stellar mass histograms of spec-IR and three radio-IR sub-samples. \textit{Panel C:} \zspec~histogram of spec-IR and three radio-IR sub-samples. Median values of each sample are shown in dashed vertical lines and 1$\sigma$ uncertainties are shown as shaded regions with the same colours as used in the histograms. The p-value result of the K-S test (\pks) between each radio-IR sub-sample and the overall radio-IR/spec-IR sample are shown in the legend after labels. For a reference, if \pks $<$ 0.05, the two distributions are not drawn from the same distribution. }
    \label{fig:properties}
\end{figure}

\subsection{Radio Galaxy Colours}
\label{sec:color}
It has been found that radio galaxies occupy the full range of colour from blue star-forming galaxies to galaxies which host traditionally-selected RAGN that are red and quiescent, while IR detected galaxies are predominantly in the star forming region. 
Here, we apply a two-colour selection technique proposed by \citet{Williams2009} to divide the galaxies into two categories: quiescent and star-forming (SF). We adopt the rest-frame of $\mathrm{M_{NUV} - M_r}$ versus $\mathrm{M_r - M_J}$ colour-colour diagram following separation lines from \citet{Lemaux2014}. 
Specifically, the line is defined as that region within $\mathrm{M_{NUV} - M_r > 2.8(M_r - M_J) + 1.51}$ and $\mathrm{M_{NUV} - M_r > 3.75}$ at 0.55 $\leq$ z $\leq$ 1.0 and region within $\mathrm{M_{NUV} - M_r > 2.8(M_r - M_J) + 1.36}$ and $\mathrm{M_{NUV} - M_r > 3.6}$ at 1.0 $<$ z $\leq$ 1.3 are considered quiescent. We then calculate a ``colour offset'' value for galaxies as their perpendicular offset to quiescent and star-forming separation lines according to their \zspec, with positive representing galaxies in the quiescent region. 

In the panel A of Figure \ref{fig:properties}, we show the colour offset histograms for the spec-IR and the three radio-IR sub-samples. The median values are marked as vertical lines and shaded by 1$\sigma$ uncertainty. 
Uncertainties on the median colour offset are given by $\sigma_{\text{NMAD}}/\sqrt{n-1}$ where $\sigma_{\text{NMAD}}$ is the normalized median of the absolute deviations \citep{Hoaglin1983} and $n$ is the number of the sample (see \citealp{Shen2019}). Throughout the paper we conservatively adopt $\sigma_{\text{NMAD}}/\sqrt{n-1}$ as the formal uncertainty. 
It appears that the median colour offset of SFGs is clearly separated from others, with their colour offset histogram extending to lower values. 
We employed the Kolmogorov-Smirnov statistic (K-S) test and the resultant \pks~value\footnote{In this paper, we adopt the K-S test and the resultant \pks~value that if \pks $>$ 0.05, we cannot reject the hypothesis that the two distributions are drawn from the same distribution. Otherwise, we say the probability of drawing from the same distribution is very small. } to test whether the colour offset distributions of three sub-samples and the overall radio-IR/spec-IR samples are consistent with being draw from the same distribution. 
The \pks values on the SFG sample are 0.03 and $\approx 0$, which confirm that SFGs do not share the same colour offset distribution as the radio-IR and spec-IR samples. 
For low- and high-\fAGN, the median values largely overlapped with each other and that of the spec-IR sample, with differences between high-\fAGN~and spec-IR distribution found in the K-S test. The AGN sub-samples are, on average, closer to the quiescent region, indicating their stellar populations are older than that of SFGs. To test this point, we take the age of the main stellar population in the galaxy (``sfh.age\_main'' parameter) estimated in CIGALE which represents the time since the onset of star formation and then following a delayed SFH. The average age of SFGs is $\sim$2 Gyr, while that of the two AGN sub-samples are $\sim$3 Gyr. 

Overall, MIR detected galaxies dominate in the star forming region (i.e., colour offset $<$ 0). 26\% of spec-IR are quiescent galaxies, compared to 40\% for the overall spectroscopy galaxies in the same redshift and stellar mass range in the ORELSE survey \citep{Lemaux2018b}. The fractions of quiescent galaxies are even smaller, $\sim$10\% for the three radio-IR sub-samples. AGN are predominantly hosted by galaxies in the star-forming region, which in line with studies that show IR detected RAGN are not hosted by red and quiescent galaxies (e.g., \citealp{Magliocchetti2018}) and other studies of obscured AGN (e.g., \citealp{Chang2017a}). 

\subsection{Stellar Mass}
The stellar mass histograms are shown in panel B of Figure \ref{fig:properties}. We find that all radio-IR sub-samples are hosted by more massive galaxies than the overall spec-IR sample at $\gtrsim 3\sigma$ level. This result is consistent with studies showing that the probability for a galaxy to be a radio source increases with increasing stellar mass (e.g., \citealp{Ledlow1996}). 
Furthermore, it seems that low- and high-\fAGN~hosts are more massive than SFGs. Specifically, we find that high-\fAGN~hosts skew toward higher stellar masses among the three radio sub-samples. We apply K-S tests to confirm this difference. The \pks s between low-/high-\fAGN~and spec-IR are small, implying they are not drawn from the same distribution.  The K-S tests is not conclusive between the SFG sub-sample and the overall spec-IR sample. 
Again, as compared to traditionally-selected RAGN hosts, at z$\sim$1 and the same stellar mass limit, the median stellar mass is found to be $10^{11.16}$ \citep{Shen2017}. Neither high- nor low-\fAGN~reach this median value. 
It is likely that AGN selected in this study are not hosted by the same type of galaxy hosting traditionally-selected RAGN that are massive and dominate the quiescent region. 

\subsection{Redshift}
In panel C of Figure \ref{fig:properties}, we show the redshift distribution of the three radio-IR sub-samples, as well as the spec-IR sample. The median \zspec s largely overlap with each other. 
None of the K-S tests conclusively confirm any difference between the sub-samples and the overall radio-IR and spec-IR sample. Therefore, we can safely ignore redshift-driven evolutionary effects in this study. 

\subsection{Environments}
\label{sec:properties_env}

\begin{figure*}
    \includegraphics[width=\textwidth]{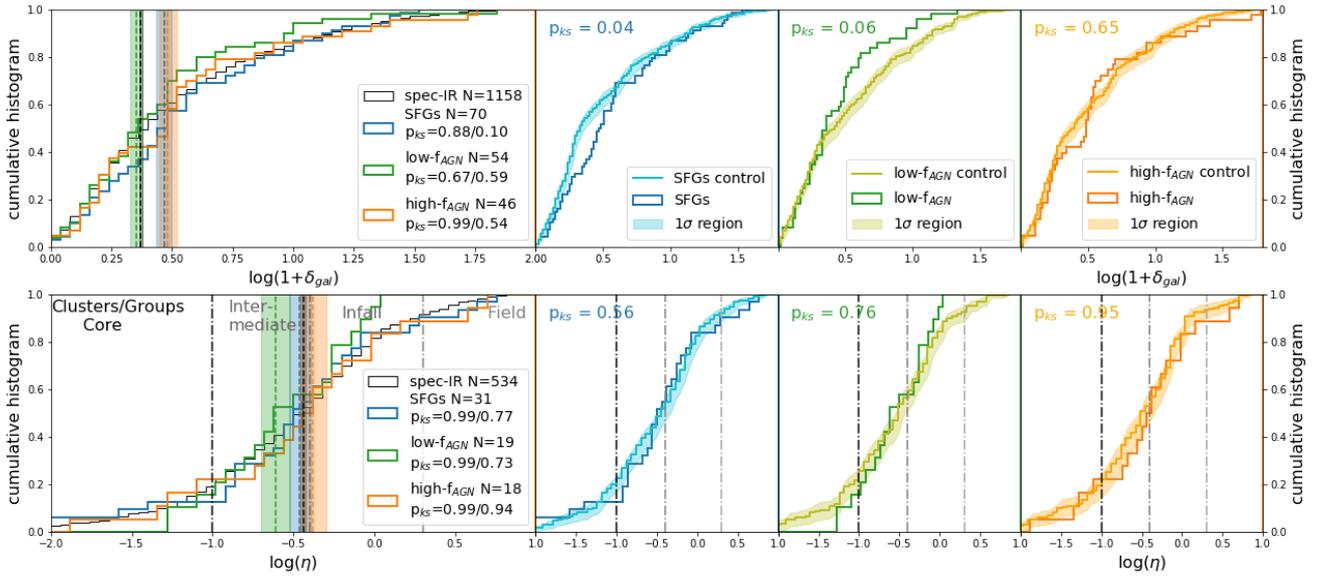}
    \caption{\small Local and global environment Cumulative Distribution Functions (CDFs) of the three radio-IR sub-samples and spec-IR sample.  \textit{Top left}: CDFs of \ldelta~of high-\fAGN~(orange line), low-\fAGN~(green line) and SFGs (blue line) and spec-IR (black dashed line). The median values are shown in coloured dashed vertical line with shaded region corresponding to 1$\sigma$ uncertainties. The number of galaxies in each sample in both plots is displayed after their label, as well as the \pks~values of the K-S test between each sub-sample and the overall radio-IR/spec-IR sample. \textit{Top right panels}: from left to right corresponding to comparisons of local \ldelta~CDFs of SFGs, low-, high-\fAGN~sub-sample. The lines with shaded regions are median values and the 1$\sigma$ spread (i.e., the 16\% and 84\% values) of the 100 control samples. The mode of \pks~are presented in each panel. \textit{Bottom}: CDFs of \leta~of high-\fAGN~(orange line), low-\fAGN~(green line), SFGs (blue line) and spec-IR (black dashed line), limited to galaxies having $R_{proj}/R_{200} < 3$ and $|\delta v/\sigma_v| < 3$. This cut limit galaxies to be associated with an individual cluster/group. For a reference, three dashed lines of constant \leta~are displayed to separate between clusters/groups core ($\eta \leq 0.1$), intermediate ($0.1 < \eta \leq 0.4$), infall regions ($0.4 < \eta \leq 2$) and field region ($\eta > 2$). \textit{Bottom right panels}: from left to right, CDFs of \leta~of SFGs, low- and high-\fAGN, and median values and 1$\sigma$ spreads (i.e., the 16\% and 84\% values) of the 100 control samples. Three lines of constant \leta=0.1, 0.4, 2.0 are displayed as well. }
    \label{fig:env}
\end{figure*}

RAGN are preferentially found in the cores of galaxy clusters and locally overdense environments both in the local universe (e.g., \citealp{Miller2002, Best2004}) and at high redshifts ($z>0.5$, e.g. \citealp{Best2014, Magliocchetti2004, Lindsay2014}). 
However, RAGN detected in MIR are preferentially found in the field at $z \leq 1.2$ \citep{Magliocchetti2018}. 
On the other hand, SFGs are broadly distributed in terms of local overdensity \citep{Tomczak2019}.   
Therefore, it is imperative to explore the role of both the clusters/groups and local environment on the three radio-IR sub-samples.

We plot the Cumulative Distribution Functions (CDFs) of the \ldelta~of each radio-IR sub-sample, as well as the overall trend of the spec-IR sample in the left top panel of Figure \ref{fig:env}. 
We find that the median \ldelta~of the low-\fAGN~galaxies overlaps with the median value of the spec-IR sample, being lower than that of SFGs and high-\fAGN~galaxies. The latter two sub-samples share similar median values. 
To confirm a local environmental effect, rather than those due to galaxy type and stellar mass that can bias the analyses of environments, we draw 100 control samples from the spec-IR sample, excluding radio galaxies, matching each radio-IR sub-sample on colour and stellar mass. 
In brief, a control sample is identified using a 3D matching algorithm, following the \citet{Shen2017} method, which ensures that the distributions of M$_*$ and two rest-frame colours ($\mathrm{M_{NUV} - M_{r}}$, $\mathrm{M_{r} - M_{J}}$) of the control sample match closely to those of the parent radio sub-samples. K-S tests are run on the \leta~and \ldelta~distributions of the parent and control sample. 
In order to explore the full breadth of possible outcomes for this comparison, we perform 100 iterations of control sampling and K-S testing. The mode of the \pks~is obtained by binning \pks s into 10 bins and returning the bin with the most \pks. 
In the right panels of Figure \ref{fig:env}, we show the local overdensity CDF of each radio-IR sub-sample and median CDF of its control samples, along with a 1$\sigma$ (i.e., the 16\% and 84\% values) shaded region. From left to right panels correspond to SFG, low-, and high-\fAGN~comparison. The modes of the \pks s are shown in the panels. 
The SFG-control sample appears slightly different from the SFG population in the less dense regions, which is confirmed by the small mode of the \pks. 
This result indicates that SFGs have a preference to be in locally denser environments than its control sample. 
The CDF of the low-\fAGN~galaxies shows an enhancement in the intermediate density region compared to its control sample. The mode of \pks~is 0.06 which is close to the significance threshold of the K-S test. This result might hint that the local environmental preference of low-\fAGN~is the intermediate density region. Even though there are small differences in the high-\fAGN~panels, the K-S tests are not conclusive. We will further interpret these results when combining the results from the global environment.

With the recently updated clusters/groups catalogs in the ORELSE fields \citep{Hung2019}, we are able to measure the global environment in a more consistent way. We found that 40\% of the SFGs are in the cluster/group environment (i.e., \leta~$\leq$ 2) versus 32\% of galaxies in the low-\fAGN~galaxies and 44\% in the high-\fAGN. 
To obtain a more quantitative look at the environment distribution within clusters/groups, we plot the \leta~CDFs of each sub-sample as well as the overall trend of spec-IR sample in the bottom panels of Figure \ref{fig:env}. Note that we adopt the cut of $R_{proj}/R_{200} \leq 3$ and $|\delta v/\sigma_v| \leq 3$ to include galaxies that are clearly associated with an individual cluster/group. The number of galaxies in each sub-sample after applying this cut are shown in the legend. 
It appears that the median values of the three radio-IR cluster/group sub-samples and the spec-IR cluster/group sample largely overlap, with their median values in the intermediate and infall regions. 
We find that low-\fAGN, on average, have marginally lower \leta~values than that of SFGs and high-\fAGN, which seems like an opposite behavior as shown in local overdensity distributions. 
It might due to the difference of sample selections, i.e., galaxies associated with clusters/groups or field galaxies. The median \leta~of the overall SFG, low-, and high-\fAGN~sub-samples are 0.71, 1.11 and 0.73, respectively, which is consistent with the local overdensity preferences. Note that a larger \leta~means the galaxy resides further away from the cluster/group cores. The relatively large median \leta~value of the overall low-\fAGN~sub-sample suggests that there are more field galaxies. 
We further test if the local density preferences is due to galaxies associated with clusters/groups or in field. We find that the median \ldelta~of clusters/groups galaxies in the low-\fAGN~sub-samples is 0.61 $\pm$ 0.06 which is lower than that of other two sub-samples (0.68 $\pm$ 0.06 for high-\fAGN~and 0.79 $\pm$ 0.04 for SFGs), while the median \ldelta~values of field galaxies in the three sub-samples overlaps within their uncertainties. These might suggest that it may be clusters/groups galaxies in low-\fAGN~that possibly assert themselves to give the local environmental preference. 
None of the K-S tests on the distribution of \leta~of the three radio-IR clusters/groups sub-samples and the radio-IR/spec-IR clusters/groups sample are conclusive as shown in the legends after the labels. 

We again run the control sample comparison as applied in the analyses of local environment to investigate more carefully the cluster/group preference. Note that here we draw control samples from the spec-IR sample within $R_{proj}/R_{200} \leq 3$ and $|\delta v/\sigma_v| \leq 3$, matching galaxies within this phase space in each radio-IR sub-sample. 
As shown in the bottom right three panels of Figure \ref{fig:env}, the \leta~distributions of the controls largely overlap with their parent radio-IR populations. These results suggest that radio-IR sub-samples do not have any preference in terms of cluster/group environments compared to the overall spec-IR sample and colour-stellar mass-matched control samples. 
However, we note that the number of galaxies in each sub-sample is small, only $\sim$ 20 galaxies in each, which might obscure differences between these populations. 

Combining the results from the local and global environments, the SFG population marginally tends to be located in locally dense regions, corresponding to the intermediate/infall regions of clusters/groups, with such local environmental preference persisting when comparing to its colour-stellar mass control sample. The properties of the SFG population in this paper are largely consistent with the overall SFG population found in the LSS \citep{Shen2017}, in terms of colour, local and global environment, though a higher stellar mass on average is seen in the LSS. 
We note that although the two studies use the same stellar mass cut and are at comparable radio depth, different classification methods are adopted, which might lead to the small differences in the stellar mass. 

The low-\fAGN~sub-sample tends to be located in locally intermediate density regions, corresponding to the intermediate/infall regions of clusters/groups, with such local environmental preference persisting when comparing to its colour-stellar mass control sample. The high-\fAGN~sub-sample shares the same environmental preference both locally and globally as their colour-stellar mass control samples. Both AGN sub-samples avoid the core of clusters/groups where traditionally-selected RAGN are preferentially to be located \citep{Hickox2009, Shen2017}. 
These results are consistent with other studies of radio galaxies. 
\citet{Bardelli2010} found that RAGN hosted by SFGs and pure SFGs at $0 < z < 1$ do not show significantly different environmental distributions compared to control samples with similar infrared colour and specific SFR, while traditionally-selected RAGN show a significant preference to be in denser regions. \citet{Magliocchetti2018} showed that RAGN that emit at MIR have a significant preference to be in the less dense field-like environment, compared to RAGN that are not detected in MIR. 

\begin{figure}
\centering
    \includegraphics[width=\columnwidth]{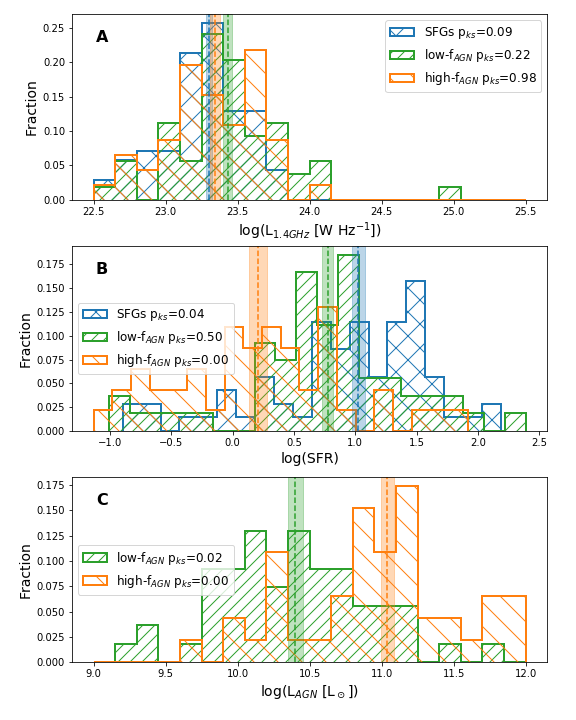}
    \caption{\textit{Panel A:} \Lradio~histograms of three Radio-IR sub-samples. \textit{Panel B:} \lsfr~histograms of three Radio-IR sub-samples. \textit{Panel C:} \LAGN~histograms of three Radio-IR sub-samples. Median and 1$\sigma$ uncertainty values of each sample are shown in dashed vertical line and shaded region with the same colours as used in the histograms. Uncertainties on \Lradio, \lsfr, \LAGN~of sample are the $\sigma_{\text{NMAD}}/\sqrt{n-1}$. The p-value result of the K-S test (\pks) between the three Radio-IR sub-samples to the overall Radio-IR are shown in each panel after the label. In the bottom panel, the \pks s are the K-S test results between the high-/low-\fAGN sub-samples to the combined high- and low-\fAGN sample. For a reference, if the \pks < 0.05, the two distributions are not drawn from the same distribution. }
    \label{fig:properties2}
\end{figure}

\subsection{Radio Luminosity}
\label{sec:radio_lum}
In studies of RAGN, radio luminosity is seen to strongly correlate with the black hole mass (M$_\text{BH}$) and anti-correlate with Eddington ratio (\rEdd = L$_{bol}$/\LEdd), using a variety of AGN classification methods (e.g. \citealt{Laor2000, Ho2002, McLure2004, Sikora2007, Chiaberge2011, Sikora2013, Ishibashi2014}). On the other hand, thanks to the well established far-infrared-radio correlation (\citealp{Condon1992, Kennicutt1998, Yun2001}), radio luminosity is a good indicator of the SFR for SFGs \citep{Bell2003, Hopkins2003}. 

In panel A of Figure \ref{fig:properties2}, we show the histograms of radio luminosity (\Lradio) of the three radio-IR sub-samples. 
The three histograms are statistically identical, with the low-\fAGN~sub-sample having, on average, slightly higher radio power than the other two sub-samples.
There are 7 galaxies having \Lradio $\leq 10^{23.8}$ W Hz$^{-1}$, 6 in low-\fAGN~and 1 in high-\fAGN. 
This radio luminosity threshold is the typical radio luminosity cut for RAGN at z $\sim$ 1 by other studies (e.g., \citealp{Hickox2009, Shen2017, Shen2019}). 
Note that we apply this radio power threshold when selecting the SFGs population but not in the low- and high-\fAGN~sub-samples. 

\subsection{Star Formation Rate}
\label{sec:sfr}
In panel B of Figure \ref{fig:properties2}, we show the histograms of star formation rate (SFR) derived in CIGALE of the three radio-IR sub-samples. 
It appears that SFGs have, on average, the highest SFR, then the low-\fAGN, while the high-\fAGN~extend to lower values. 
The median radio luminosity for our SFGs is log(\Lradio) = 23.30, corresponding to SFR = 66.38 M$_\odot$yr$^{-ˆ'1}$, using the star formation rate formula from 1.4GHz from \citet{Bell2003} and converting Salpter IMF to Charibier IMF by multiplying by a factor of 0.6. However the median SFR derived from CIGALE is 10.56 M$_\odot$yr$^{-1}$. There are two reason for this discrepancy. One reason is that a rapid, strong episode of star formation could produce long-lived stars that contaminate the bands tracing star formation for hundreds of Myr afterwards. This contribution is corrected for CIGALE, which in turn lowers the SFR derived in this method \citep{Boquien2014, Boquien2016}. An other reason is that SFR estimators calibrated on long timescales are valid for galaxies with long, gradual star formation episodes but not for starbursting galaxies. Therefore, assuming a single exponential decaying SFH might underestimate the SFR for these SFGs as they are in a starbursting episode.

\subsection{AGN Power}
\label{sec:agn_lum}
In panel C of Figure \ref{fig:properties2}, we show the histograms of AGN luminosity (\LAGN) of the high- and low-\fAGN~sub-samples. 
We find a clear offset in both median values and their histograms. 
The K-S tests on the \LAGN~distribution of high-\fAGN~and low-\fAGN~is $<$ 0.02, confirming the difference in \LAGN~distributions. 
In the radio luminosity comparison, we found a marginally higher, on average, radio luminosity for low-\fAGN. 
These results seem inconsistent with the general picture of AGN luminosity increasing with increasing radio luminosity \citep{Sikora2013, Ishibashi2014}, since the bulk of the energy is released from AGN in a kinetic form via radio jets \citep{Churazov2005}. However, these studies are mainly focused on the powerful RAGN \Lradio $> 10^{25}$, much higher than the radio luminosity of our sample. 

In fact, studies of radio sources suggest the origins of radio luminosity from AGN accretion \citep{Zakamska2016, White2015, White2017} and/or star-forming activities (e.g., \citealp{Kimball2011, Condon2013}). In order to see if a combination of mechanisms could explain our result, we apply a simple test. 
From the AGN side, following a radio luminosity and X-ray luminosity correlation with a slope of 1.2 $\pm$ 0.15 \citep{Panessa2015} calibrated for AGN at high Eddington ratios (i.e., $(L_\text{AGN}/L_\text{Edd}) > 10^{-3}$), we obtain $\Delta \text{log}L_{1.4GHz}$ = $1.2 \Delta \text{log}L_{X-ray}$. We then assume that AGN bolometric luminosity $L_\text{AGN} \propto L_\text{X-ray}$ with a constant correction factor \citep{Brightman2017}; thus, $\Delta \text{log}L_\text{1.4GHz} = 1.2 \Delta \text{log}L_\text{AGN}$. We apply this correlation to the observed difference in \LAGN, assuming that for Type-2 AGN \LAGN~mostly contributes to the total IR range. The median \LAGN~of high-\fAGN~galaxies is 0.6 dex higher than that of low-\fAGN galaxies, corresponding to $\sim$0.5 dex higher in radio luminosity. From the SFR side, the median SFR of high-\fAGN~galaxies is 0.5 dex lower than that of low-\fAGN galaxies, following the radio luminosity - SFR correlation \citep{Bell2003}, corresponding to the same difference in radio luminosity. 
Thus, the confluence of the radio emission generated by the two mechanisms could lead to the similar median values of radio luminosity of high- and low-\fAGN sub-samples. Note that it is not necessary to include different weights on the contribution of AGN versus SF, as these have been included in the differences of the observed \LAGN~and SFR, and the conversion between these two contributions and radio luminosity are on the same order. 

As a summary for the analyses of this point, the overall radio-IR galaxies are situated in the dusty star-forming colour-colour region, at the massive end of the overall spec-IR sample. Specifically, the SFG population comprises the most active star forming galaxies. The high-\fAGN~sub-sample are hosted by the most massive galaxies, have the least star formation activity, and occupy the high end of AGN luminosity. The host galaxies of the low-\fAGN~sub-sample have stellar mass and star formation rates in between these two sub-samples. We do not find significant differences in radio luminosity of these sub-samples, which could be due to conversation of radio emission from the combined level of AGN and SF activity. 
Combining these results in the two AGN sub-samples, high-\fAGN~galaxies, on average, have higher stellar mass, lower SFR and higher \LAGN. If an evolution sequence can be made following an increase of stellar mass, low-\fAGN~would evolve to high-\fAGN, with their AGN becoming a larger contribution to the internal energy injection to the host galaxy and their SFR subsequently being quenched, which is consistent with the AGN-driven quenching scenario (i.e, \citealp{Fabian2012, Page2012, Lemaux2014}). However, AGN on their ramp down phase, when their AGN contributions and luminosity decrease, would also fall in our low-\fAGN~sub-sample, Therefore, we should have two galaxy populations in the low-\fAGN~sub-sample, one having high SFR and one with their SFR quenched due to the AGN feedback. 
To further test this co-evolution scenario, we mainly focus on these two sub-samples in the next half of this paper, depending on AGN and SF activity in their own sub-samples. 

In the comparison of environments, SFGs tend to preferentially reside in locally dense environments, and the low-\fAGN~sub-sample shows a marginal preference for locally intermediate density region, comparing to their control samples. However, the high-\fAGN~sub-sample does not show any difference in local environment as compared to its control samples. All sub-samples share the same global preference as their control samples which reside in the intermediate/infall regions of clusters/groups and avoid the core region. 
If there is an evolutionary trend between high- and low-\fAGN, their similar environmental preference suggests that their environments, on average, do not change on bulk during this evolution. We will continue the analyses of environment in Section \ref{sec:phase} and discussion in Section \ref{sec:discussion_env}.

\section{The Correlation of SFR and AGN Luminosity}
\label{sec:SFR_AGN}

\begin{figure*}
    \includegraphics[width=\textwidth]{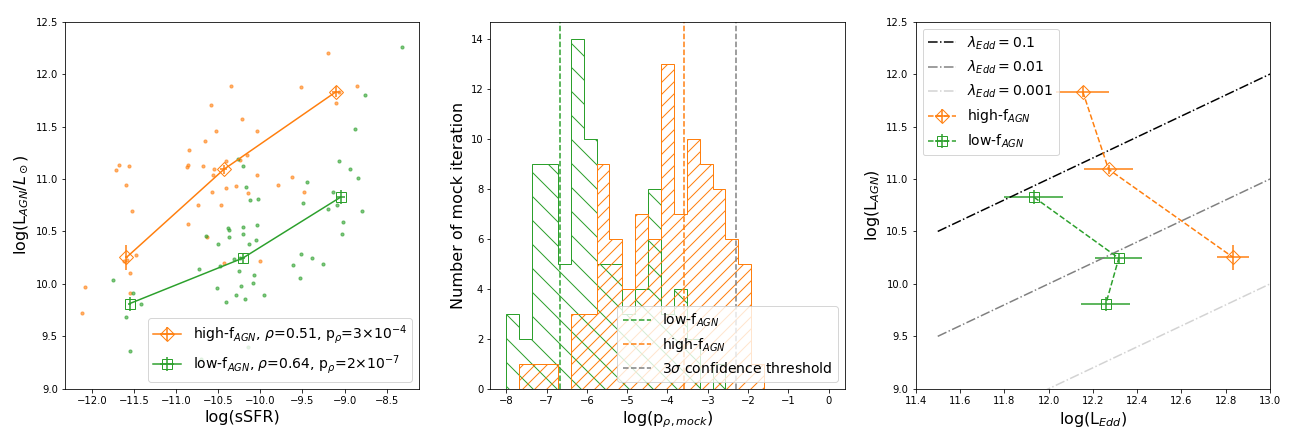}
    \caption{\small \textit{Left}: Relation between \LAGN~ and \lssfr~of high-\fAGN~(orange) and low-\fAGN~(green) sub-samples. The orange and green open data points and lines show median values in each \lssfr~bin for galaxies in the high-/low-\fAGN~sub-sample, respectively. Strong and significant positive correlations between \LAGN~and \lssfr~are seen in both samples. Note that SFGs are not shown in here since they show no or very minimal AGN activity. \textit{Middle}: Distribution of the logarithmic value of p$_{\rho}$ between \LAGN~and \lssfr. This distribution is calculated from 100 Monte Carlo realizations in which we randomly assign mock\fAGN~and corresponding mock-\LAGN~and -\lssfr~values drawn from the mock catalogues. Orange and green vertical dashed lines are log(p$_{\rho}$) of real sub-samples, as shown in the left panel. For reference, a log(p$_{\rho}$)<-2.3 corresponds to a correlation between the two parameters estimated at $\geq 3 \sigma$ confidence level, indicated by the grey vertical dashed line. \textit{Right}: Median of \LAGN~and \LEdd~for high-\fAGN~(orange) and low-\fAGN~(green) in each \lssfr~bin shown in the left panel. Three grey dashed lines indicate \rEdd = 0.1, 0.01, 0.001 from top to bottom, respectively. }
    \label{fig:sSFR_Lagn}
\end{figure*}

Many studies have demonstrated that AGN activity (e.g., AGN bolometric luminosity) correlates with SF activity especially for luminous AGNs (e.g., \citealp{Netzer2009, Gurkan2015}). As we show in Figure \ref{fig:properties2}, it seems like AGN luminosity is anti-correlated with SFR for the two AGN sub-samples. However, as we mentioned at the end of the last section, each sub-sample may have a mix of AGN in different states, which might lead to an overall anti-correlation relation. 
Here, we attempt to investigate the correlation of AGN and SF in their own sub-samples. 
We plot \LAGN~and \lssfr~of high-\fAGN~(orange dots) and low-\fAGN~(green dots) sub-samples in the left panel of Figure \ref{fig:sSFR_Lagn}. The green and orange open data points and lines show median values in each \lssfr~bin for galaxies in the high-/low-\fAGN~sub-sample, with $\sigma_{NMAD}/\sqrt{n-1}$ as their errors. 
Note that SFGs are not shown in here. Some of galaxies in the SFG population show low AGN activity. However, due to the difficulty of constraining low AGN contributions, with an overestimation up to a factor of 2 for \fAGN $<$ 0.1 (\citealp{Ciesla2015}, see also Figure \ref{tab:CIGALE} in Appendix \ref{app:CIGALE_compare}), we do not include them in this analyses. 

We find that \LAGN~increases with increasing \lssfr~for both the low-\fAGN~and high-\fAGN~samples. 
To further formalize this result, we calculated the Spearman rank correlation coefficient, $\rho$, as well the significance of the rejection of the null hypothesis of no correlation, p$_\rho$. We adopt p$_\rho <$ 0.05 as significant threshold and p$_\rho <$ 0.005 as $\geq3\sigma$ significant threshold. We find $\rho$ = 0.51 and 0.64 between \LAGN~and \lssfr~for the high- and low-\fAGN~sub-samples, respectively. The small p$_\rho$ values (p$_\rho$ = 0.001 and 0.006 for the high- and low-\fAGN~sub-samples, respectively) show the rejection of the null hypothesis of lack of correlation between the two variables at the $\gg 3\sigma$ level. 

To incorporate the uncertainty of \fAGN~and the corresponding uncertainties in \LAGN~and \lssfr~values derived from CIGALE, a Monte-Carlo simulation was run in the following manner. In CIGALE, a mock SED was generated for each galaxy by varying each photometric point randomly sampled from a Gaussian distribution of observed photometric flux and its corresponding error. We then evaluate the best-fit SED based on the mock photometries and derive the mock results. We run mock SEDs 100 times for each galaxy to obtain 100 mock-\fAGN, mock-\LAGN, and mock-sSFR values. In each mock iteration, galaxies are then split into sub-samples given their mock-\fAGN, and a Spearman test is run on the mock-sSFR and mock-\LAGN~values. In the middle panel of Figure \ref{fig:sSFR_Lagn}, the histograms of p$_\rho$ of low- and high-\fAGN~mock realizations are shown. The correlation remains at $\gg 3\sigma$ significant level in almost all for low- and high-\fAGN~mock realizations. 
As confirmed by this test, it appears the positive correlation of the \LAGN~and \lssfr~relation observed in the high- and low-\fAGN~sub-samples is a real feature of galaxies in the radio-IR that have an AGN component. If we combine the high- and low-\fAGN~sub-samples, the p$_\rho$ is 0.06, indicating no correlation would be found if the galaxies were not divided on \fAGN.
These positive significant correlations strongly suggest a co-evolution of AGN and SF in AGN found in the radio-IR galaxies.

\subsection{Estimating the States of the AGN}
\label{sec:AGNstates}

Studies from observations and simulations have found that AGN duty cycles and lifetimes depend on Eddington ratio (\rEdd) and black hole mass (M$_\text{BH}$). Specifically, lifetimes increase as decreasing Eddington ratio and decreasing black hole mass (e.g., \citealp{Martini2004, Hopkins2009}). In fact, the general picture broadly showstopper  that the lifetime at high Eddington ratio (\rEdd $>$ 0.1) is $\sim$ 100Myr and increases rapidly at lower Eddington ratio to $\sim$0.5 - 1 Gyr for \rEdd $>$ 0.01 and $\sim$1 - 5 Gyr for \rEdd $>$ 0.001. In other word, \rEdd~can be used as a tracer of the AGN state and lifetime. 

We assume, again, that for Type-2 AGN \LAGN~mostly contributes to the total IR range. For a central object with M$_\text{BH}$, the Eddington luminosity is given by \citet{Shapiro1983}:
\begin{equation} 
    \label{eq:LEdd}
    L_\text{Edd} = 3 \times 10^4 (M_\text{BH}/M_\odot) L_\odot.
\end{equation}
We adopt the correlation of M$_\text{BH}$ with bulge mass M$_\text{bulge}$ following equation 10 in \citet{Kormendy2013} to estimate M$_\text{BH}$ and assume a bulge-to-total stellar mass ratio of 0.3 for galaxies \lm $<$ 11 and 0.4 for \lm $\geq$ 11 found in \citet{Lang2014}, which were derived from a star forming galaxy population selected in UVJ colour-colour diagram at $0.5 < z < 2.5$. We confirm this ratio on average (mean) equal to 0.3 with a standard
deviation of 0.3 using 142 galaxies in the ORELSE survey that occupy the star-forming region of the NUVrJ colour-colour diagram and have stellar mass in a range of 8.3$<$\lm$<$11 and median at 9.7 \citep{Pelliccia2019}\footnote{The bulge-to-total stellar mass ratio is computed using the galaxy's total and spheroid component fluxes measured with the SExtractor (v2.19.5, \citealp{Bertin1996, Bertin2011}), which fitted each galaxy's HST/ACS F814W image with a Sersic+exponential disk profile model convolved with the local PSF.}. 
We estimated the median \LEdd~for each bin in the low- and high-\fAGN~sub-samples binning by log(sSFR) as shown in the left panel of Figure \ref{fig:sSFR_Lagn}. In the right panel of Figure \ref{fig:sSFR_Lagn}, we plot the estimated median log(\LEdd) versus observed log(\LAGN) for low- and high-\fAGN~bins. The uncertainty of the Eddington luminosity is estimated combining the error of stellar mass in each bin and the 0.29 dex intrinsic scatter in equation 10 in \citet{Kormendy2013}. Note that we do not include the error from the bulge-to-total stellar mass ratio, though including the uncertainly of bulge-to-total stellar mass ratio $\pm$ 0.1 would not affect our result. 

\begin{figure*}
    \includegraphics[width=\textwidth]{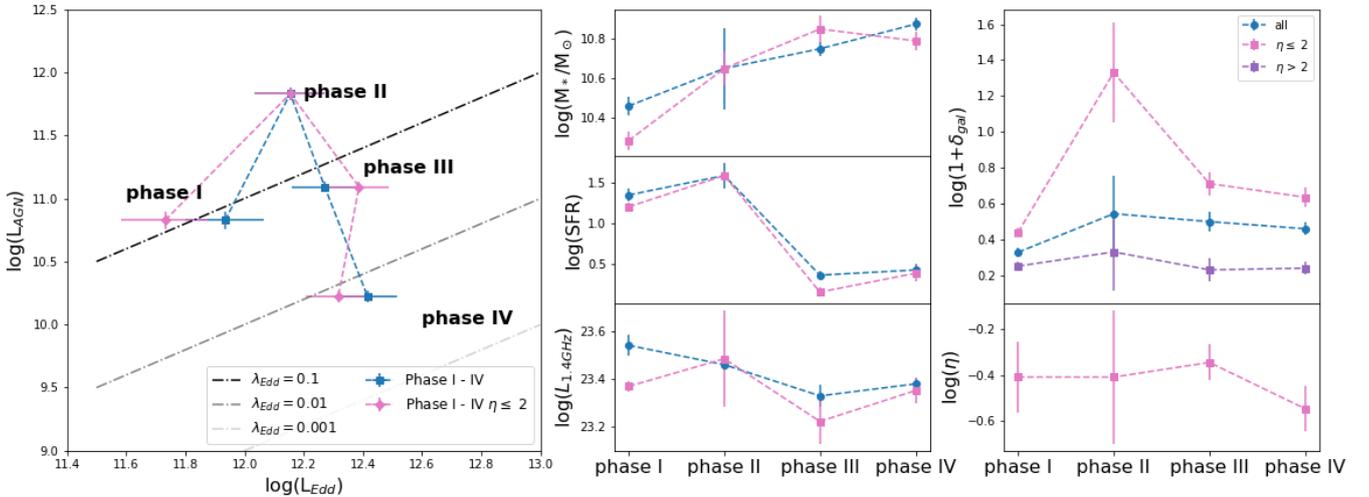}
    \caption{\textit{Left}: Median \LAGN~versus \LEdd~of all galaxies in each phase proposed in Section \ref{sec:phase} are shown in blue diamonds and those within cluster/group region ($\eta \leq$ 2) in pink squares for four phases. \textit{Middle (from top to bottom):} Median stellar mass, star formation rate, radio luminosity of all galaxies in four phases and those within cluster/group regions. \textit{Right top}: Median values of local overdensity of all galaxies in four phases and those within cluster/group regions and in the field. \textit{Right bottom}: Median values of global environment of galaxies within cluster/group regions of the four proposed phases.}
    \label{fig:phase}
\end{figure*}

\begin{table*}
	\caption{Median properties of galaxies in Phase I - IV overall and within cluster/group region ($\eta \leq$2)}
	\label{tab:phase}
	\begin{threeparttable}
	    \begin{tabular}{c|ccccccccc} 
        \hline
        Phase (all) & Num. & log(\LAGN) & log(\LEdd) & \rEdd & \lm & \lsfr & \Lradio & \ldelta & \leta \\
        \hline
        \hline
        Phase I & 16 & 10.83$\pm$0.07 & 11.93$\pm$0.13 & 0.08 & 10.46$\pm$0.05 & 1.34$\pm$0.08 & 23.54$\pm$0.04 & 0.33$\pm$0.03 & 1.20$\pm$0.15\\
        Phase II & 5 & 11.83$\pm$0.05 & 12.15$\pm$0.12 & 0.37 & 10.65$\pm$0.21 & 1.59$\pm$0.15 & 23.45$\pm$0.01 & 0.55$\pm$0.21 & 0.005$\pm$0.66\\
        Phase III & 29 & 11.09$\pm$0.04 & 12.27$\pm$0.11 & 0.04 & 10.75$\pm$0.04 & 0.36$\pm$0.05 & 23.33$\pm$0.04 & 0.50$\pm$0.05 & 0.62$\pm$0.16\\
        Phase IV & 50 & 10.22$\pm$0.04 & 12.42$\pm$0.10 & 0.007 & 10.88$\pm$0.03 & 0.43$\pm$0.07 & 23.38$\pm$0.03 & 0.46$\pm$0.03 & 1.11$\pm$0.09\\
        \hline
        \hline
        Phase ($\eta \leq$ 2) & Num. & \LAGN & \LEdd & \rEdd & \lm & \lsfr & \Lradio & \ldelta & \leta \\
        \hline
        \hline
        Phase I & 4 &  10.83$\pm$0.07 & 11.73$\pm$0.15 & 0.12 & 10.29$\pm$0.05 & 1.20$\pm$0.06 & 23.37$\pm$0.02 & 0.44$\pm$0.02 & -0.41$\pm$0.15\\
        Phase II & 3 & 11.83$\pm$0.05 & 12.15$\pm$0.12 & 0.47 & 10.65$\pm$0.08 & 1.59$\pm$0.06 & 23.48$\pm$0.20 & 1.33$\pm$0.3 & -0.41$\pm$0.29\\
        Phase III & 13 & 11.09$\pm$0.04 & 12.39$\pm$0.10 & 0.05 & 10.85$\pm$0.07 & 0.15$\pm$0.06 & 23.22$\pm$0.10 & 0.71$\pm$0.07 & -0.34$\pm$0.08\\
        Phase IV & 18 & 10.22$\pm$0.04 & 12.32$\pm$0.11 & 0.008 & 10.79$\pm$0.05 & 0.39$\pm$0.10 & 23.35$\pm$0.05 & 0.64$\pm$0.05 & -0.55$\pm$0.10\\
        \hline
        \end{tabular}
    \end{threeparttable}
\end{table*}

In the right panel of Figure \ref{fig:sSFR_Lagn}, three dashed lines from top to bottom indicate \rEdd~$=$ 0.1, 0.01, \& 0.001, respectively, where $\lambda_\text{Edd} = L_\text{AGN} / L_\text{Edd}$. We see that the high-\fAGN~sub-sample occupies the large space from \rEdd~= 0.5 to 0.003, while the highest \rEdd~is 0.08 in the low-\fAGN~sub-sample. There are only 5 galaxies in the high-\fAGN~high-sSFR bin at high \rEdd~versus 29 in the high-\fAGN~middle-sSFR bin and 16 in the low-\fAGN~high-sSFR bin, two bins at lower \rEdd~region. These relative differences in number of galaxies is consistent with the lifetime of AGN, which is much shorter at high \rEdd. 
In addition, the number of galaxies in the low-\fAGN~low-sSFR bin is small (5 galaxies). The reason for this small sample size is likely due to the threshold of the observed L$_\text{AGN}$. 
For less massive galaxies that host less massive black holes and therefore lower Eddington luminosities to have the same state of AGN that are hosted by more massive galaxies (i.e., same \rEdd), they would be detected at lower AGN luminosity. However, due to the detection limit and the SED fitting threshold of lower AGN luminosity objects, we are not complete for these lower AGN luminosity objects hosted by less massive galaxies. This is also seen in our data, where the stellar mass of hosts in the low-\fAGN~low-sSFR bin is $10^{10.74}M_\odot$ which is lower than that in the high-\fAGN~low-sSFR bin ($10^{11.11}M_\odot$). Therefore, we argue that as AGN luminosity decreases and their SF is quenched, they also fall out of the sample. We will further discuss these post-AGN and post-star forming galaxies in Section \ref{sec:discussion_env}. 

\subsection{A Co-evolution Scenario}
\label{sec:phase}

To explore the idea that the two AGN sub-samples might contain AGN in different states (i.e., ramp up or ramp down phase), and to confirm whether the lower SFR in the high-\fAGN~sub-sample is due to AGN-driven quenching, we create additional bins according to their median \rEdd and SFR: designating galaxies in the low-\fAGN~high-sSFR bin to be ``Phase I'', galaxies in high-\fAGN~high-sSFR bin to be ``Phase II'', galaxies in high-\fAGN~middle-sSFR bin to be ``Phase III'', and the remaining middle-sSFR bin and low-sSFR bins to be ``Phase IV'' as they are all predominantly found in low \rEdd~phase space. 

In the left panel of figure \ref{fig:phase}, the median \LAGN~and \LEdd~of all galaxies in each phase are shown in blue diamonds and only those within the cluster/group region ($\eta \leq$ 2) in pink squares. 
Phase II has, on average, \rEdd $\sim$0.4, Phase I and Phase III have, on average, \rEdd $\sim$0.08 and $\sim$0.04 respectively, while Phase IV has, on average, \rEdd $\sim$0.007, which implies that AGN in Phase II are in the most active state, while those in Phase IV are the least active. Adopting AGN lifetimes from \citet{Hickox2012}, from Phase I to Phase III is approximately 100 $\sim$ 500 Myr at their median estimated M$_\text{BH}$ $\sim 10^{7.5} M_\odot$. 
The lifetime of AGN at Phase II is $<$ 100 Myr and the overall timescale from Phase I to Phase IV is $>$1 Gyr. 

In the right two columns of Figure \ref{fig:phase}, we show the median stellar mass, star formation rate, radio luminosity, and environmental preference of all galaxies in the four phases and those within cluster/group regions. To investigate the co-evolution of AGN and SFR in cluster/group environments, we focus here mostly on the difference in properties for those galaxies in the cluster/group region, i.e. $\eta \leq$ 2. The number of galaxies in each phase and the median values of these properties are shown in Table \ref{tab:phase} for all galaxies and galaxies in clusters/groups. 

\subsubsection{Stellar Mass and SFR Evolution}
\label{sec:phases_M_SFR}
In the middle-top panel of Figure \ref{fig:phase}, we show the median stellar mass from Phase I to IV for all galaxies (blue) and galaxies in cluster/group regions (pink). A significant increase as galaxies move through these phases is shown, regardless of overall or in the cluster/group environments, which supports a co-evolution scenario. 

In the middle-middle panel, the median values of SFR are shown. It is clear that SFR drops rapidly from Phase II to III and remains constant to Phase IV. This trend suggests that from Phase I/II to later Phase III, SF must shut down within a relatively short timescale ($\sim$ 100Myr in \citealp{Hopkins2009}). Again, such a trend is seen in both the total and cluster/group galaxy sample. 
This result strongly suggests AGN-driven quenching in SFGs, in line with other observations in AGN hosted by SFGs selected in X-ray and IR bands (e.g. \citealp{Page2012, Lemaux2014}) and residual of AGN or other feedback signatures found in post-starbust galaxies (e.g., \citealp{Yan2006, Kocevski2009b, Rumbaugh2017, Lemaux2017}). In addition, this is supported by the theory that AGN are capable of removing most of the gas from their host galaxies by winds and outflows, leaving a galaxy unable to form stars (e.g., \citealp{Springel2005, Hopkins2007, Somerville2008}). 
We apply a simple test to see whether a rapid quenching is required. Given the decrease of SFR by $\sim$1 dex, and an exponential decay star formation history (SFH), SFR $\propto e^{−t/\tau}$, this would correspond to $t/\tau \approx$ 2.3. For a typical short $\tau$ model, i.e., $\tau$=0.2 Gyr, it would take $\sim$500 Myr. For a longer $\tau$ model, i.e., $\tau$=2 Gyr, the corresponding timescale is 4 Gyr. However, the timescale from phase I to phase III is no more than several 500 Myr; therefore, a short $\tau$ model is required to reach the observed decrease in SFR. In Section \ref{sec:model}, we will further discuss this test, combining with proper AGN models, and quantify the quenching timescale. 

Furthermore, observations in the local universe have suggested a significant delay between the starburst phase and the peak of nuclear optical AGN activity (e.g., \citealp{Davies2007, Schawinski2009, Wild2010, Yesuf2014}). Indeed, in realistic hydrodynamic models, AGN luminosities do not peak at the same time as the gas density on small scales, but rather $\sim 10^8$ yr later when most of the diffuse gas has been exhausted by star formation and/or expelled by stellar feedback \citep{Hopkins2012}.

\subsubsection{Radio Luminosity Evolution}
\label{sec:phases_radio}
In the middle-bottom panel of Figure \ref{fig:phase}, the median radio luminosity appears flat in AGN through this evolution scenario. There is a slight decrease from Phase I to Phase III ($\sim$0.1 dex in between each phases), while such decrease is not shown to be statistically significant in the cluster/group sample, likely due to the small number of galaxies in each phase. 
The decrease of radio power could be due to the quenching of star formation or lower radio jet power or both. 
In Section \ref{sec:radio_lum}, we have shown that radio emission is originated from the combined AGN and SF activities. However, here we see that theSFR from Phase I to Phase III decreases by $\sim$1 dex, corresponding to a 1 dex decrease in \Lradio. The median \LAGN~from Phase I to Phase III increases by 0.2 dex, corresponding to a $\sim$0.2 dex increase in \Lradio. If we simply add these two quantities together, we are not able to obtain the difference in the radio luminosity ($\sim$0.2 dex). The reason for this finding could be due to the different timescales associated with AGN activity and star formation as well as a delay between the two \citep{Wild2010}. In addition, in a study of radio galaxies at $z \sim 1$ down to \Lradio $\sim 10^{22.5}$ W Hz$^{-1}$, \citet{White2017} found that in the early stages of AGN-SF co-evolution, radio luminosity is dominated by emission associated with SF because of the SF has not been quenched and the radio jets have not been fuelled. In the later stages, radio luminosity is taken over by AGN when the SF is being quenched and progressively lower accretion rates allow radio jets to form. Therefore, we can explain the \Lradio, on average, due to the the contribution by AGN and SF, whereas the same explanation failed once we split AGN to different stages which probe shorter timescales. 

\subsubsection{Environmental Evolution}
\label{sec:phases_env}
In the right column of Figure \ref{fig:phase}, we show the environments associated with the four phases. In the right-top panel of Figure \ref{fig:phase}, the median values of local overdensity of all galaxies (blue), galaxies in clusters/groups ($\eta \leq$ 2 in pink) and galaxies in the field ($\eta >$ 2 in purple) in each phase. 
From Phase I to Phase IV,  there is a general movement of the galaxy populations to move to more locally dense environments and, in the case of phase IV, possibly to globally denser environments. These galaxies, at least those in the clusters and groups, still predominantly remain in the outskirts/infall region of their parent structure, not in the core, for the entirety of this process. 
Indeed, galaxies in Phase III show a higher local density than Phase I at 3$\sigma$ for all galaxies and 3.7$\sigma$ for galaxies in clusters/groups (\leta $\leq$ 2), but no significant offset is seen in the field. 
We see that galaxies in Phase II show a higher local density in all three samples. However, the errors of this phase are large, mainly due to small sample size, so the differences are not significant. 

In the right-bottom panel of Figure \ref{fig:phase}, we plot the median values of global environment for galaxies in clusters/groups across the phases. All phases tend to preferentially reside in the infall region, with Phase IV being slightly closer to the core region at 2$\sigma$ significance compared to Phase III. However, due to the small number of galaxies and the associated large errors in \leta, we do not observe any significant change in terms of global environmental preference. 

After Phase IV, galaxies might have completely quenched their SF and have lower AGN luminosity, so that they fall out of the sample completely. 
In order to compensate for this relative lack of galaxies in our samples, we introduce the study of \citet{Lemaux2018b}, which uses $\sim$ 5000 spectroscopically-confirmed galaxies in the 15 ORELSE fields including the eight fields in this study and at the same redshift range. 
Galaxies with the stellar mass of the average Phase IV galaxies are right at the interface of the intermediate and high \ldelta~ in the high stellar mass sample where galaxies are found predominately quenched at all redshift probed by ORELSE (see figure 5 in \citealp{Lemaux2018b}), while such phenomenon is not seen for galaxies in other phases. 
This is consistent with our hypothesis that these galaxies are almost certainly on their way to becoming quiescent. 
These galaxies would then tend to be in the cluster/group core regions as massive quiescent galaxies. We will further discuss the effect of environment in Section \ref{sec:discussion_env}. 

\section{Discussion}
\label{sec:discussion}

In Section \ref{sec:SFR_AGN}, we found a positive relationship between \LAGN~and \lssfr, binned by \lssfr, for both high-\fAGN~and low-\fAGN~galaxies at the $\gg3\sigma$ level, which strongly suggests an evolutionary trend between AGN and SF. The relation persists when the uncertainty in \fAGN~estimated in CIGALE is included. 
We then calculated \LEdd~and sort galaxies further into Phase I to IV according to the median Eddington ratio and SFR of each bin. We found that during this proposed evolutionary scenario, AGN activities ramp up and down, as stellar mass increases constantly and SFR decreases dramatically at the AGN activity ramp down phase, which strongly suggests an AGN-driven feedback on SF in their hosts. 
In Section \ref{sec:model}, we further discuss the possible co-evolution of AGN and SF using a simple toy model including an AGN history and different SF histories (SFHs). 
Furthermore, during this scenario, their environments of the galaxies do not change appreciably, with the galaxies remaining mostly in the relatively local dense regions that are equivalent to the intermediate/infall regions of clusters/groups. 
In Section \ref{sec:discussion_env}, we will discuss the role of environment in this scenario. 

\subsection{A Simple Model to Contextualize AGN Quenching of Star-formation histories}
\label{sec:model}

\begin{figure}
\centering
    \includegraphics[width=\columnwidth]{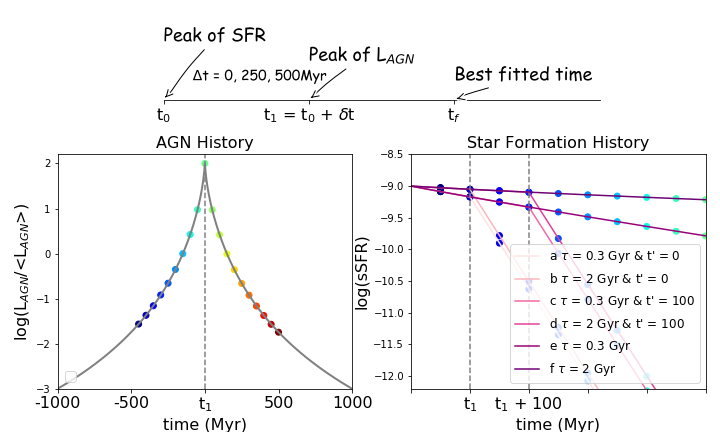}
    \caption{\textit{Top:} A schematic illustration of toy model presented in Section \ref{sec:model}. \textit{Bottom left:} The AGN history relative to the the long-term average luminosity~$<$\LAGN$>$ over 2 Gyr. The luminosity distribution varies from 10$^{-3}$ to $10^{2}$ relative to~$<$\LAGN$>$. \textit{Bottom right:} six star formation history models following an exponential decay SFH with long-$\tau$ = 2 Gyr and short-$\tau$ = 0.3 Gyr. Once quenching happens, we adopt a rapid-$\tau$ = 50 Myr model. In scenario a/b, this rapid quenching happens immediately after the peak of \LAGN~at t$_0$; in scenario c/d rapid quenching is delayed t' = 100 Myr after peak of \LAGN~at t$_0$; while in scenario e/f, no quenching happens followed by normal short-/long-$\tau$ SFH models. The coloured points are marked at each 50 Myr time stamp. }
    \label{fig:model}
\end{figure}

In this section, we create a simple toy model to contextualize the timescale of each bin in Figure \ref{fig:sSFR_Lagn} and whether SF is quenched during the evolution. 
In addition, since we found that during the transition between phase I to IV, galaxies in cluster/group environments and in the field share the same M$_*$, SFR, AGN and radio luminosity trends, we use the overall radio-IR sample as the observed galaxy population. 
In our model galaxies start their star formation at t$_0$. 
The AGN is ignited and peaks at t$_1$ = t$_0$ + $\delta$t, where $\delta$t = 0, 250, or 500 Myr, which corresponds to what we define as scenarios SI, SII, SIII, respectively. Once the AGN luminosity reaches its peak, it is allowed to affect star formation, in terms of immediate rapid quenching, delayed rapid quenching, or no effect. These scenarios are represented by the quantity t' which is allowed to be 0, 100, or $\infty$. The illustration of this toy model is shown in the top panel of Figure \ref{fig:model}. We now present details of the AGN history (AGNH) and the star formation histories (SFHs) adopted in this toy model. 

First, we adopt the AGN luminosity distribution by integrating the Eddington ratio distributions from \citet{Hickox2014}, originally from \citet{Hopkins2009}, which were derived using a range of AGN observations. The AGN luminosity distribution is modeled by a Schechter function, defined as a power law with an exponential cutoff near the Eddington limit: 
\begin{equation} \label{eq:AGNfunction}
\frac{dt}{d\text{log(L)}} = t_0\ (\frac{\text{L}}{\text{L}_{\text{peak}}})^{-\alpha}\ \text{exp}(\text{L}/\text{L}_{\text{cut}})
\end{equation}
 
$\alpha$ = 0.2 and L$_{peak}$ = 100 $<$\LAGN$>$ are adopted from \citet{Hickox2014}, in order to include a wide diversity in AGN accretion rates and luminosity distributions that are obtained from theoretical models and observational studies of different galaxy populations. For $\alpha$ = 0.2 and L$_{cut}$ = 100$<$\LAGN$>$ where $<$\LAGN$>$ is the long-term average luminosity, \LAGN~can reach to a lower limit $\sim 10^{-5}$ $<$\LAGN$>$. The distribution is normalized by scaling t$_0$, such that the integral over 10$^{-3}$ to $10^{2}$ relative to $<$\LAGN$>$ is equal to 1. 
This range is chosen based on the $\sim$ 3 dex \LAGN~range in our sample on one end, and $<$\LEdd$>$ $\sim$ $10^2<$\LAGN$>$ on the other end. 
We assume that the average AGNH is monotonic and symmetric. Therefore, a half AGN history (L(t)) can be obtained by inverting and integrating eq. \ref{eq:AGNfunction}. We adopt a 2 Gyr AGN duty cycle, consistent with general predictions in \citet{Hopkins2009} that cover the time spent from $10^{-3}$ $<$\LAGN$>$ through peak AGN luminosity and then back to $10^{-3}$ $<$\LAGN$>$. This AGN luminosity distribution as a function of time is shown as the grey line in Figure \ref{fig:model} bottom left panel, where t$_1$ is the peak of AGN luminosity. We define the exact value of $<$\LAGN$>$ below. 

For stellar population models we assume exponential decay SFHs of the form SFR $\propto e^{-t/\tau}$. We adopt three $\tau$: long-$\tau$ = 2 Gyr, short-$\tau$ = 0.3 Gyr, and rapid-$\tau$ = 50 Myr. It has been found that all star-forming galaxies in a given stellar mass bin and in low- to intermediate-density environments in the ORELSE survey are best fitted to $\tau$ = 2 Gyr, for SFHs characterized be a single $\tau$ model \citep{Tomczak2019}. 
We assume that a simulated galaxy follows a normal SFH with either a long-$\tau$ = 2 Gyr or a short-$\tau$ = 0.3 Gyr. Once quenching happens due to AGN, if it indeed occurs in the model selected, the SFH follows a rapid-$\tau$ = 50 Myr. In scenario a/b, this rapid quenching happens immediately after the peak of \LAGN~at t$_1$; in scenario c/d rapid quenching is delayed t' = 100 Myr after peak of \LAGN~at t$_0$, motivated by a lag time between the peak of AGN luminosity and SFR (see references in Section \ref{sec:phases_M_SFR}), while in scenario e/f, no quenching happens and the original SFH is allowed to progress unimpeded.  
The amount of stellar mass generated through in situ star formation in each time step is added to the previous stellar mass of the galaxy by integrating the star formation rate from t=0 to that step and subtracting the amount of stellar mass lost over that same time period resulting from stellar evolution \citep{Moster2013}. 
In the bottom-right panel of Figure \ref{fig:model}, the sSFR for the six SFHs models are shown with initial log(SFR) = 1.5 and log($M_*/M_\odot$) = 10.5 (see below for these choices). The coloured points are marked at each 50 Myr time stamp. 

\begin{figure}
    \includegraphics[width=\columnwidth]{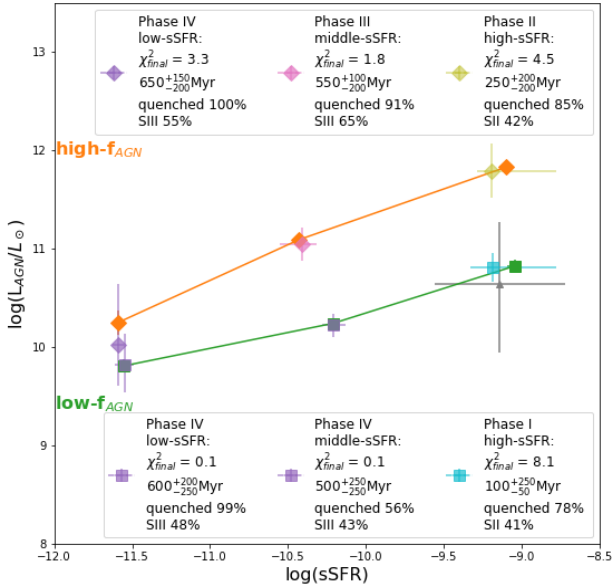}
    \caption{The median of the best-fitted log(sSFR) and log(\LAGN) and the observed median log(sSFR) and log(\LAGN), the same as shown in the left panel of Figure \ref{fig:sSFR_Lagn}. The uncertainties of the best-fitted log(sSFR) and log(\LAGN) are the 16th/84th percentiles. The diamond symbols are observed and simulated data points for high-\fAGN~and the squares are low-\fAGN, colour coded according their phases proposed in Section \ref{sec:phase}. The median of initial sSFR and the assigned $<$\LAGN$>$ are shown as a grey triangle with error bar on the 16th/84th percentiles. The $\chi^2_{final}$ reported in the legend is calculated from the median of the best-fitted to the median and error of the observed data points. The median of best-fitted time with uncertainty and the preferred SFH and AGNH scenario with their percentage of simulated galaxies are listed in the legend. }
    \label{fig:simulation_result}
\end{figure}

In each realization, a simulated galaxy is assigned an initial SFR and M$_*$ as well as a $<$\LAGN$>$, and is then allowed to evolve following all combinations of AGNH and SFH models. 
Since we do not know the initial SFR and stellar mass of galaxy before the AGN ignited, we use the distribution of SFR and M$_*$ of SFGs in the high-sSFR bin (log(sFFR) $>$ -9.5).  Specifically, the initial state of this simulated galaxy is randomly sampled from a Gaussian distribution with the mean and variances of SFR and M$_*$ of this bin ($<$log(SFR)$>$ = 1.46 $\pm$ 0.33 and $<$\lm$>$ = 10.48 $\pm$ 0.31). 
To represent the possible \LAGN~of the observed galaxy population, we use the combined distribution of the estimated \LAGN~of the two observed AGN sub-samples (low- and high-\fAGN). 
Specifically, we randomly assign a $<$\LAGN$>$ to this simulated galaxy sampled from a Gaussian distribution with the mean and variance of \LAGN ($<$\LAGN$>$ = 10.68 $\pm$ 0.64). 
The median and 16th/84th percentiles of initial sSFR and the assigned $<$\LAGN$>$ are shown as a grey triangle with error bar in Figure \ref{fig:simulation_result}. 
Note that the assigned $<$\LAGN$>$ might not be the initial state of the simulated galaxy. According to the three AGN scenarios, at t$_0$, \LAGN~can be 100$<$\LAGN$>$, 0.22$<$\LAGN$>$ or 0.02$<$\LAGN$>$ in SI, SII or SIII, respectively. 
Each combination of AGNH and SFH model produces a distinct set of predictions for the evolution of the sSFR and \LAGN~of galaxies. We then find the best-fitted model and timescale to observations by minimizing a residual $\chi^2$, as defined as: 
\begin{subequations}\label{eq:chi2}

$\chi^2 = (\frac{\Delta \text{log(sSFR})}{\sigma \text{log(sSFR)}})^2 + (\frac{\Delta \text{log(L}_{\text{AGN}})}{\sigma \text{log}(\text{L}_{\text{AGN}})})^2$ \\

$\Delta \text{log(sSFR}) = \text{log}(\frac{\text{sSFR}_{\text{obs}}}{\text{sSFR}_{\text{model}}})$ \\

$\Delta \text{log}(\text{L}_{\text{AGN}}) = \text{log}(\frac{\text{L}_{\text{AGN, obs}}}{\text{L}_{\text{AGN, model}}})$\\
\end{subequations}

where $\sigma$log(sSFR) and $\sigma$log(\LAGN) represent the uncertainties on the observed log(sSFR) and \LAGN~respectively.
In addition, we fit a separation line in the log(\LAGN)-log(sSFR)~phase space, where above this line is the high-\fAGN~model and below is the low-\fAGN~model. During the process of finding the best scenario for the observed points, we only consider models that lie in the same category. 

In order to explore the full breadth of possible outcomes for this comparison, we run these simulations for 1000 realizations. The median of the best-fitted log(sSFR) and log(\LAGN) are shown in Figure \ref{fig:simulation_result} for both the high- and low-\fAGN~galaxies, plotted in comparison to the observed median log(sSFR) and log(\LAGN). The uncertainties of the best fits are the 16th/84th percentiles. The colours indicate their phases proposed in Section \ref{sec:phase} and symbols according their sub-samples. 
The median of best-fitted t$_f$ and 16th/84th percentiles are listed in the legend after their labels. We define the preferred scenario by counting best-fitted simulated galaxies in each AGNH and SFH scenario, separately. The most frequent one is the preferred AGNH/SFH, as shown in the legend, along with their percentage in the best-fitted simulated galaxies. 
In addition, we calculate a $\chi^2_{final}$ from the median of the best-fitted simulated models to the median and error of the observed data points in each bin. 

It appears that the simulated points for the high-sSFR bins are offset to the left, with their $\chi^2_{final}$s larger than other bins. This is mostly driven by the time for \LAGN~to match the observations, during which SFR must decline due to the adopted exponential decay SFH. 
This might suggest that our toy model is not as appropriate here as it is for bins that have evolved after a longer time, mostly due to lack of information on the initial states of these phases. At longer times, such bias is migrated by simulating a large variety of initial states. 

The best-fitted $t_f$s are, in general, consistent with the co-evolution scenario in which the best-fitted $t_f$s, on average, increase from Phase I to Phase IV. However, we do see that the best-fitted $t_f$ of Phase III and the low-\fAGN~middle-sSFR bin which is in Phase IV overlap within their uncertainties. 
In the right panel of Figure \ref{fig:sSFR_Lagn}, the median \rEdd~of low-\fAGN~middle-sSFR bin is shown to be the highest among the three bins included in Phase IV, indicating that they are closer to the peak of AGN and closer to Phase III. 

Furthermore, we find that in all cases, a rapid quenching $\tau$ model is preferred. 
For the two high-sSFR bins, though quenching scenarios are picked up more often, we find that the median $\chi^2$ of no quench models ($\tilde{\chi^2} \sim$ 7 for high-\fAGN~and $\tilde{\chi^2} \sim$ 3 for low-\fAGN) is much lower than that of quenched models ($\tilde{\chi^2} \sim$ 48 for high-\fAGN~and $\tilde{\chi^2} \sim$ 32 for low-\fAGN). Combined with the relatively short time evolved in the simulation, we are not able to draw a definite conclusion on quenching/no quenching of these bins. 
For the high-\fAGN~middle-sSFR bins, quenching models are picked up $\sim$90\% of the time, suggesting rapid quenching is required. 
As for the low-\fAGN~middle-sSFR bin, quenching and no quenching models are picked close to half and half in percentage, 
indicating that both scenarios are equally preferred. These results suggest that the simulated galaxies can follow all AGNH/SFHs combinations and are best-fitted in a wide $t_f$ range to fit the observed point of this bin. In fact, complex types of galaxies might fall in this bin with their moderate SFR ($\sim$40 M$_\odot$ yr$^{-1}$) and low \LAGN. One possibility is a galaxy with its AGN in its ramp up phase and its SFR slowly decreasing because of gas exhaustion. It is also possible that galaxies with low luminosity AGN are not powerful enough to quench SF in their hosts. 
For the high-sSFR bins, in almost all cases, quenching models are picked, which strongly suggests that quenching must happen and is due to the AGN. 
We note that here we do not separate immediate rapid quenching (a/b) or delay rapid quenching (c/d) scenarios, because the 100 Myr delay time is short; thus, the difference between having a delay or not is buried in the time of the best-fitted $t_f$ and its uncertainties. 

As for the quenching timescale, in Phase III, simulated galaxies are dominated by scenario SIII with AGN peak at $t_1$ = 500 Myr and the quenching timescale of 50 Myr. For low-sSFR bins of Phase IV, a longer quenching timescale is suggested to be 100$\sim$200 Myr. We also found that all galaxies in Phase III are in the star-forming colour-colour region and have low SFR = 2.3 M$_\odot$ yr$^{-1}$ on average, suggesting that these galaxies are on their way to quiescence. On the contrary, 65\% galaxies in the combined low-sSFR bins of Phase IV are in the quiescent colour-colour region with their log(sSFR) $\leq$ -11, a signature of quiescent galaxies. 
Such a quenching timescale broadly agrees with studies of the observed fraction of FIR-detected host galaxies in X-ray up to $z \sim 3$ for massive galaxies ($\leq10^{10}M_\odot$) ($\sim$100 Myr, \citealp{Mullaney2012, Page2012, Rosario2012}). 
In addition, \citet{Hickox2012} found that powerful starbursts preferentially reside in the same environment as AGN at $z \sim 2$ with their estimated lifetime of $\sim$100 Myr. They argued that powerful starbursts and AGN occur in the same systems, and AGN could be responsible for the short lifetime. 
In simulations, \citet{Mancuso2017} estimated \Lradio~by applying the global star formation rate functions at different redshift and evolutionary tracks for the history of star formation and BH accretion in an individual galaxy based on the continuity equation and the abundance matching technique. 
They recovered well the observed radio luminosity function in the redshift range of $0.5 < z < 4.5$ with \Lradio$ > 10^{23.5}$ at $z \sim 1$. Their model is highly consistent with our toy model but adopted a wider  rapid quenching star formation $\tau$ model  ($\tau \leq 10$ Myr). 

\subsection{The Effect of Environments}
\label{sec:discussion_env}
As we found in Section \ref{sec:properties_env}, the AGN sub-samples do not show significant differences in their environmental preference either locally or globally, when comparing to the overall spec-IR sample and their control samples. They tend to be located in the intermediate/infall regions of cluster/group environments. This preference persists during the proposed co-evolution scenario (see Section \ref{sec:phase}). In our toy model, we found that the timescale of this co-evolution is up to $\sim$ 700 Myr, which might hint that these galaxies are not infalling radially into a cluster/group, but rather have tangential orbits in their clusters/groups. 

Simulations of galaxies in clusters/groups have shown that galaxies are not directly falling into the core region of the cluster/group once they are captured by the cluster/group. Indeed, galaxies may orbit beyond their host cluster virial radius, with their typical orbit timescale of several Gyr before infalling. A short orbiting timescale ($\sim$0.5 Gyr) is suggested at higher redshift $z \sim 1$ \citep{Muzzin2014}. 
The tangential orbital motion of galaxies is also supported by the ``delay-and-rapid'' quenching scenario recently studied in the ORELSE survey. Specifically, environmental-driven quenching efficiencies and associated timescales are constrained in the ORELSE survey using $\sim$ 5000 spectroscopically-confirmed galaxies in the redshift range $0.55 \leq z \leq 1.4$ and a semi-empirical model that generated mock galaxies unaffected by environmentally-specific processes \citep{Lemaux2018b}. The average timescale for SFGs in the field to be accreted by a cluster/group and become quiescent is 2.4 $\pm$ 0.3 Gyr for galaxies with log(M$_*$/M$_\odot$) $> 10^{10.45}$ and 3.3 $\pm$ 0.3 Gyr for lower stellar mass galaxies. Meanwhile, the average timescale of star-forming galaxies to be quenched is $t_{quench}$ = 1.1 $\pm$ 0.3 Gyr using the same data set \citep{Tomczak2019}. The difference between these two timescales is the $t_{delay}$, the time between accretion and the inception of the quenching process.  Such a long delay time is in excess of the average time required for galaxies to move from the outskirt of a cluster to its core region \citep{Wetzel2013, Muzzin2014}. Therefore, orbits other than purely radial are required, such as a ``splashback'' population characterized by galaxies on their second passage of the cluster (e.g., \citealp{Chang2018}). 

However, the average $t_{quench}$ = 1.1 $\pm$ 0.3 Gyr seems difficult to link to the short quenching timescale found in Section \ref{sec:model}. One explanation could be due to the fact that \citet{Tomczak2019} defined quiescent galaxies by the colour-colour diagram, while in this study AGN sub-samples are hosted by galaxies in the star-forming region in the colour-colour diagram. Even at the last phase, 34\% of AGN hosts are in the star-forming region but we argue in section \ref{sec:phases_env} that they would be on their way toward quiescence. Another possibility could be due to the difference in stellar mass. AGN hosts are dominated by massive galaxies, corresponding to the intermediate and high stellar mass bins in \citet{Tomczak2019}. Meanwhile, studies have shown only feedback from luminous AGN hosted by massive galaxies (M $> 10^{11}$ M$_\odot$) can contribute to the mass-quenching of galaxies, while other contributions are needed for lower stellar mass galaxies (e.g., \citealp{Bongiorno2016}). Nevertheless, \citet{Tomczak2019} probed the general SFG population, while our AGN selected in radio galaxies may undergo more vigorous starbursts which could also lead to more rapid gas removal. 

Furthermore, we see that cluster/group galaxies in phase IV tend to be closer to the cluster/group core regions, which might hint that as galaxies move towards the core, their AGN and SF activities might decline. 
However, our sample is not complete at low AGN luminosity and low SFR, since both MIR and radio are sensitive to galaxies that either have ongoing star forming or AGN activities. 
In fact, the co-evolution timescale is broadly consistent with studies of post-starburst/K+A populations in clusters which have recently ended a burst of star formation and/or have rapidly quenched with their $\tau$ = 0.5 - 1 Gyr (e.g., \citealp{Dressler1999, Muzzin2014, Wu2014, Lemaux2017, Socolovsky2018}). 
Specifically, in the ORELSE survey, \citet{Lemaux2017} studied a true post-starburst population in these LSSs finding that it consisted of both traditionally-selected K+A galaxies and a set of K+A galaxies with [O II] emission originating from processes other than star formation, referred as ``KAIROS'' galaxies. When comparing the AGN population in Phase IV in this paper with the true K+A population from \citet{Lemaux2017}, we find that both tend to be located in the intermediate region. They also found post-starburst galaxies contain spectral features indicate a rapid cessation of SF within $\sim$ 300 Myr and possible residual AGN activity, consistent with the quenching time estimated in our toy model. Therefore, we argue that AGN, as they follow their co-evolutionary tracks, will evolve into true K+As.  Such lines of thought will be followed further when investigating the environments of AGN and K+As in the full ORELSE sample over the full redshift range. 


\section{CONCLUSIONS}
\label{sec:conclusion}

We have studied the properties of radio-IR hosts galaxies in eight fields at 0.55 $\leq z \leq$1.30 in the ORELSE survey. We select 179 radio-IR galaxies by using VLA 1.4GHz observations down to a 4$\sigma$ detection flux density limit of about 40$\mu$Jy, matched using a maximum likelihood technique to our spectroscopically-confirmed galaxies that are also detected at $\geq 1\sigma$ in Spitzer/MIPS imaging. 
We are able to constrain the AGN and star formation contributions in IR luminosity separately by using the CIGALE SED fitting. We further separated radio-IR galaxies into SFGs, low-, and high-f$_\text{AGN}$ sub-samples primary based on the fraction of AGN contribution in total IR luminosity (f$_\text{AGN}$) estimated in the CIGALE SED fitting, with an additional radio luminosity cut for SFGs to exclude traditionally-selected RAGN. A comparison spec-IR sample was constructed using spectroscopically-confirmed galaxies detected in SPIRE/MIPS 24 $\mu$m imaging at same level as the radio-IR sample. 

Our main conclusions are the following:
\begin{itemize}

    \item The host galaxies of the overall radio-IR sample are situated in the dusty star-forming colour-colour region and dominate the massive end of the spec-IR sample. Specifically, the SFG population is the least massive and has the most active star forming galaxies among the three sub-samples. Conversely, high-f$_\text{AGN}$ are hosted by the most massive galaxies and have the least star formation activity. The host galaxies of low-f$_\text{AGN}$ have stellar masses and star formation rates in between these two populations. No significant difference between radio luminosity of these sub-samples was found, which could be due to the combined contribution of radio emission from star formation and AGN activity.  
        
	\item As for environmental preferences, the SFG population tends to preferentially be located in locally dense regions, low-\fAGN~galaxies show a marginal preference in the locally intermediate dense regions, and high-\fAGN~galaxies do not show any local environmental preference, compared to their  colour-stellar mass-matched control sample. In the cluster/group environments, none of the three sub-samples shows different environmental preference compared to their colour-stellar mass-matched control sample which are found to be located in the intermediate/infall regions of clusters/groups. 
	
	\item A positive correlation between AGN luminosity and specific star formation rate was observed in both the low- and high-f$_\text{AGN}$ sub-samples, with such correlations persisting when including the uncertainty of f$_\text{AGN}$. This result strongly suggests a co-evolution between these two activities. Based on the estimated Eddington ratio and the observed SFR, an evolution scenario was suggested following the AGN duty cycle, during which stellar mass constantly increases, whereas star formation is quenched rapidly. To contextualize this co-evolution, a toy model was created. In all cases, a rapid quenching model is preferred when matching the observed AGN luminosity and specific star formation rate, which we argue is a signature of AGN quenching.  
	
	\item The environmental preference for intermediate/infall regions in clusters/groups is similar in the two AGN sub-samples and persists during the evolutionary scenario, which suggests that as these AGN are quenching star formation in their hosts over a $\sim$ 1 Gyr timescale, they might be in an orbital motion around the clusters/groups, rather than directly infalling into the core region.
	
\end{itemize} 

Finally, as for what physical processes may be responsible for AGN and SF activities, it is likely that these activity, at least those galaxies in the groups/clusters, are induced by merging activity, since they are hosted by massive galaxies and are preferentially found in the intermediate/infall regions where merging activities are found to be common. 
Therefore, we plan to follow up our Radio-IR targets with an IFS/IFU to search for signs of ongoing or recent merging activity which might allow us to prove or disprove this hypothesis.
For future studies, we plan to continue the investigation of the coeval nature of AGN and star formation, using AGN detected in X-ray, optical, IR, and radio bands in the full ORELSE sample, to fully characterize the cycle of AGN activity and its feedback on its host galaxy. Together with the well constructed cluster/group catalog in ORELSE \citep{Hung2019} which spans a wide total dynamical mass range, we will further investigate the role of environment in the evolution of AGN, for example, as a function of cluster centric distance and parent halo mass.

\section{Acknowledgements}
\label{sec:ack}

This material is based upon work supported by the National Science Foundation under Grant No. 1411943. Part of the work presented herein is supported by NASA Grant Number NNX15AK92G. 
PFW acknowledges funding through the H2020 ERC Consolidator Grant 683184 and the support of an EACOA Fellowship from the East Asian Core Observatories Association. 
This study is based on data taken with the Karl G. Jansky Very Large Array which is operated by the National Radio Astronomy Observatory. The National Radio Astronomy Observatory is a facility of the National Science Foundation operated under cooperative agreement by Associated Universities, Inc. 
This work is based in part on observations made with the Spitzer Space Telescope, which is operated by the Jet Propulsion Laboratory, California Institute of Technology under a contract with NASA. 
SPIRE has been developed by a consortium of institutes led by Cardiff University (UK) and including Univ. Lethbridge (Canada); NAOC (China); CEA, LAM (France); IFSI, Univ. Padua (Italy); IAC (Spain); Stockholm Observatory (Sweden); Imperial College London, RAL, UCL-MSSL, UKATC, Univ. Sussex (UK); and Caltech, JPL, NHSC, Univ. Colorado (USA). This development has been supported by national funding agencies: CSA (Canada); NAOC (China); CEA, CNES, CNRS (France); ASI (Italy); MCINN (Spain); SNSB (Sweden); STFC, UKSA (UK); and NASA (USA). 
This work is based in part on data collected at the Subaru Telescope and obtained from the SMOKA, which is operated by the Astronomy Data centre, National Astronomical Observatory of Japan; observations made with the Spitzer Space Telescope, which is operated by the Jet Propulsion Laboratory, California Institute of Technology under a contract with NASA; and data collected at UKIRT which is supported by NASA and operated under an agreement among the University of Hawaii, the University of Arizona, and Lockheed Martin Advanced Technology centre; operations are enabled through the cooperation of the East Asian Observatory. When the data reported here were acquired, UKIRT was operated by the Joint Astronomy Centre on behalf of the Science and Technology Facilities Council of the U.K. 
This study is also based, in part, on observations obtained with WIRCam, a joint project of CFHT, Taiwan, Korea, Canada, France, and the Canada-France- Hawaii Telescope which is operated by the National Research Council (NRC) of Canada, the Institut National des Sciences de l'Univers of the Centre National de la Recherche Scientifique of France, and the University of Hawai'i. The scientific results reported in this article are based in part on observations made by the Chandra X-ray Observatory and data obtained from the Chandra Data Archive. 
The spectrographic data presented herein were obtained at the W.M. Keck Observatory, which is operated as a scientific partnership among the California Institute of Technology, the University of California, and the National Aeronautics and Space Administration. The Observatory was made possible by the generous financial support of the W.M. Keck Foundation. 
We wish to thank the indigenous Hawaiian community for allowing us to be guests on their sacred mountain, a privilege, without with, this work would not have been possible. We are most fortunate to be able to conduct observations from this site. 
	



\bibliographystyle{mnras}
\bibliography{reference} 



\appendix

\section{T-PHOT Simulation Test}
\label{app:tphot}
To check the performance of T-PHOT on Herschel/SPIRE images, we run a test on a simulated Herschel SPIRE 250$\mu$m image (FWHM = 25'' and 3.6''/pixel). 
We begin with a 30'$\times$30' field of view and populate it with $\sim$18,000 simulated objects. Each object is a 2D Gaussian with a randomly-assigned position, size, ellipticity, and position angle. 
Herschel/SPIRE 250$\mu$m fluxes are assigned to each of these objects in order to reproduce the observed flux density function from H-ATLAS in \citet{Valiante2016}; because of the limited depth of H\-ATLAS ($\sim$25 mJy) and the small simulated field of view, we fit a power law to the 250$\mu$m flux density function and extrapolate down to 2 mJy. Next, we convolve the simulated objects with the Herschel/SPIRE 250$\mu$m PSF. 
For the final step, noise is added to the images sampled from a Gaussian distribution and set to reproduce the background RMS $\sim$2.5 mJy of the 250$\mu$m images of XLSS005. 

\begin{figure*}
    \includegraphics[width=\textwidth]{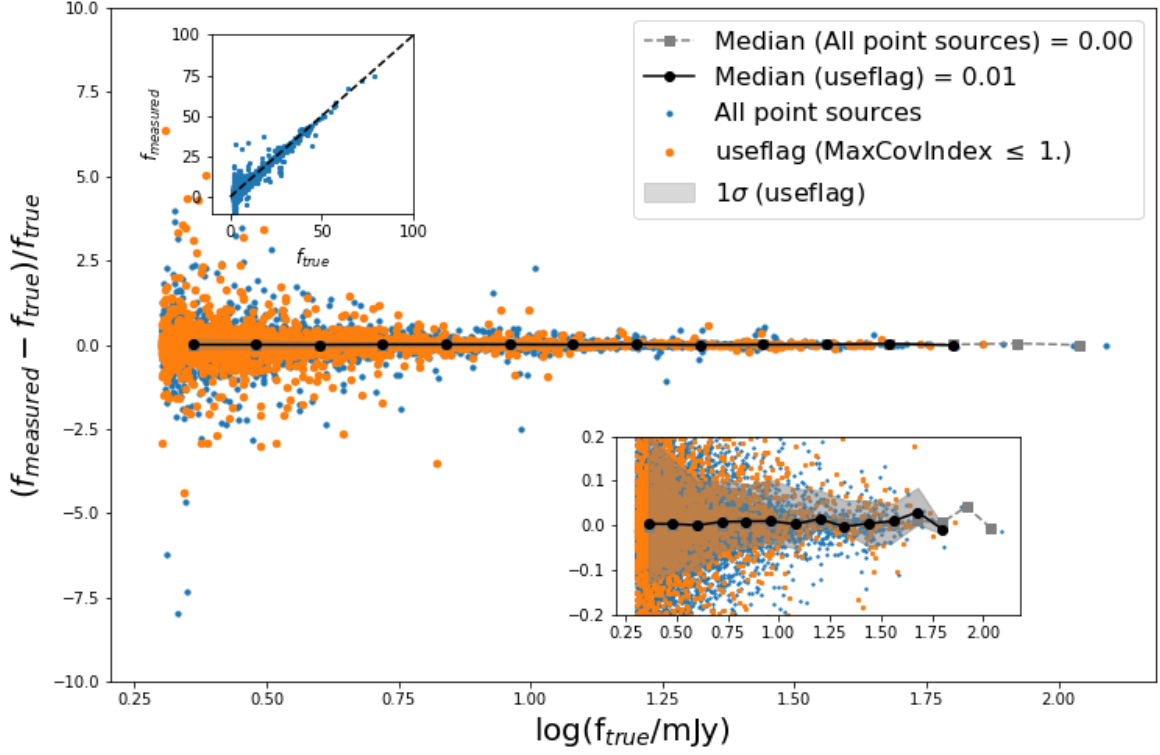}
    \caption{Accuracy of the flux determination in a FIR-like simulation (Herschel SPIRE 250$\mu$m, FWHM = 25'', 3.6''/pixel), using unresolved priors. The median of the relative flux difference ($f_{measured} - f_{true})/ f_{true}$ is close to zero, while the scatter of differences increases for low flux objects. We are confidence with the flux measured in this way down to $\sim$3 mJy where the spread of the relative flux differences is $<$ 0.1. This threshold is lower than the 1$\sigma$ depth of our Herschel/SPIRE 250 $\mu$m image shown in \ref{tab:image_depth}. }
    \label{fig:tphot_test}
\end{figure*}

For this test, T-PHOT was run using the list of simulated object positions as unresolved priors with cells-on-objects method. The results of the relative flux difference (f$_\text{measured}$ - f$_\text{true}$)/ f$_\text{true}$ versus the log of the real input flux f$_\text{true}$ is plotted in Figure \ref{fig:tphot_test}. The colour of the symbols refer to a proxy for a covariance index ``MaxCovIndex'', which is an indicator of the reliability of the fit. Large covariances often indicate a possible systematic offset in the measured flux of objects. 
The median of the relative flux differences is shown in the black solid dots, connected by the solid line, binning by simulated fluxes. The median is consistent with zero down to the faintest fluxes, independent of the simulated fluxes. This result shows that T-PHOT can recover the input fluxes of the sources with great statistical accuracy. 
The spread of the relative flux differences (16th/84th percentiles) is $\sim0.1$ for simulated objects at $\sim$3 mJy, dropping to $\sim0.05$ for simulated objects at $\sim$10 mJy. Therefore, we are confident that we can measure fluxes of objects with fluxes $>3$mJy. In addition, we found a negligible offset on the median of detection having large covariances index. Therefore, we do not include this threshold in this paper. 

\section{Uncertainty From CIGALE}
\label{app:CIGALE_compare}

\subsection{Bias From CIGALE}
\label{sec:bias}
Here we compare CIGALE derived stellar mass to FAST derived stellar mass, and CIGALE derived SFR to the SFR estimated by adding contributions from obscured and unobscured young stellar populations as traced by IR and UV emission respectively ($\text{SFR}_{\text{UV+IR}}$, \citealp{Tomczak2019}).

\begin{figure}
\centering
    \includegraphics[width=\columnwidth]{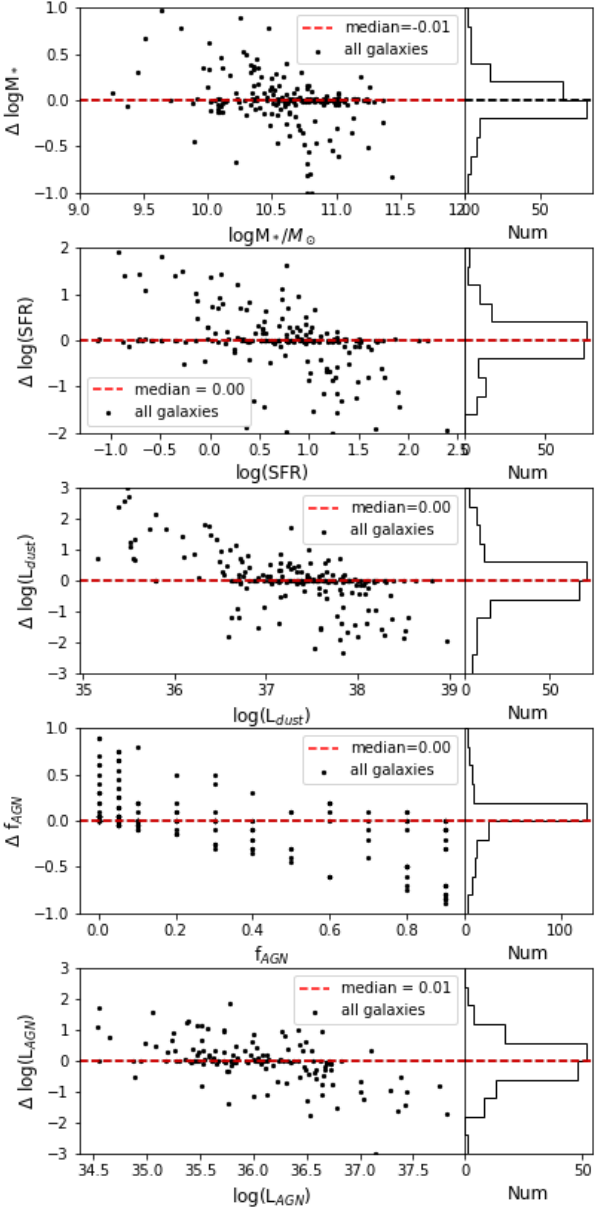}
    \caption{Comparison with physical quantities derived from CIGALE and other method for the overall radio-IR galaxies (grey dots) and those above the stellar mass cut ($M_* > 10^{10} M_\odot$) and good fitted in CIGALE ($\chi^2 \leq 10$) shown as black dots. The latter sample is the final sample analyzed in this paper. \textit{Top:} Difference of stellar mass derived from CIGALE and derived from FAST as a function of the latter quantity. Note that we use stellar mass derived from FAST in this paper, in order to perform comparisons to the overall spectroscopically-confirmed sample. The median of the difference is -0.04$^{+0.12}_{-0.21}$ with the scatter calculated from the 16th/84th percentiles. The difference is small and remain constant across the full stellar mass range studied in the paper. \textit{Bottom:} Difference of SFR derived from CIGALE and estimated by adding contributions from obscured and unobscured young stellar populations as traced by IR and UV emission respectively ($\text{SFR}_{\text{UV+IR}}$, \citealp{Tomczak2019}), as a function of the latter quantity. The median is -0.63$^{-0.22}_{-1.34}$ with the scatter calculated from the 16th/84th percentiles. The systematic higher value of $\text{SFR}_{\text{UV+IR}}$ is due to other sources of UV/dust heating (such as A or F stars) that are not tied to star formation increases in the IR luminosity and cause an overestimation of true SFR (see \citealp{Boquien2014, Boquien2016} for more discussions). }
    \label{fig:cigale_bias}
\end{figure}

In the top panel of Figure \ref{fig:cigale_bias}, we show the differences in stellar mass derived from CIGALE and those derived from FAST, for the overall radio-IR galaxies (grey dots) and those that are both above the stellar mass cut and have good fits in CIGALE ($M_* > 10^{10} M_\odot$ and $\chi^2 \leq 10$; black dots). The latter sample is the final sample analyzed in this paper. The median of difference is -0.04$^{+0.12}_{-0.21}$ with the scatter calculated from the 16th/84th percentiles. The difference is small and remains constant across the full stellar mass range that studied in the paper. Note that we use stellar masses derived from FAST in this paper in order to perform comparison to the overall spectroscopically-confirmed sample.  

In the bottom panel, we show the differences in SFR derived from CIGALE and those estimated from $\text{SFR}_{\text{UV+IR}}$. 
The median is -0.63$^{-0.22}_{-1.34}$ with the scatter calculated from the 16th/84th percentiles. A systematically higher value of $\text{SFR}_{\text{UV+IR}}$ is seen, which could due to other sources of UV/dust heating (such as A or F stars) that are not tied to star formation increases in the IR luminosity and cause an overestimation of the true SFR (see \citealp{Boquien2014, Boquien2016} for more discussions). Actually, similar trends are also seen when comparing SFR derived from FAST and Le Phare \citep{Ilbert2006} versus SFR estimated from the UV+IR method with the offset increasing at low SFR. This is a feature of SFR estimated from SED modeling, which allow SFH with low sSFR $\sim 0$, while $\text{SFR}_{\text{UV+IR}}$ will always be $>$ 0. Therefore, the offset is larger for lower SFR. 

\subsection{Comparison with Physical Quantities Derived From the Mock Catalog}
\label{sec:mockc_catalogs}
\begin{figure}
\centering
    \includegraphics[width=\columnwidth]{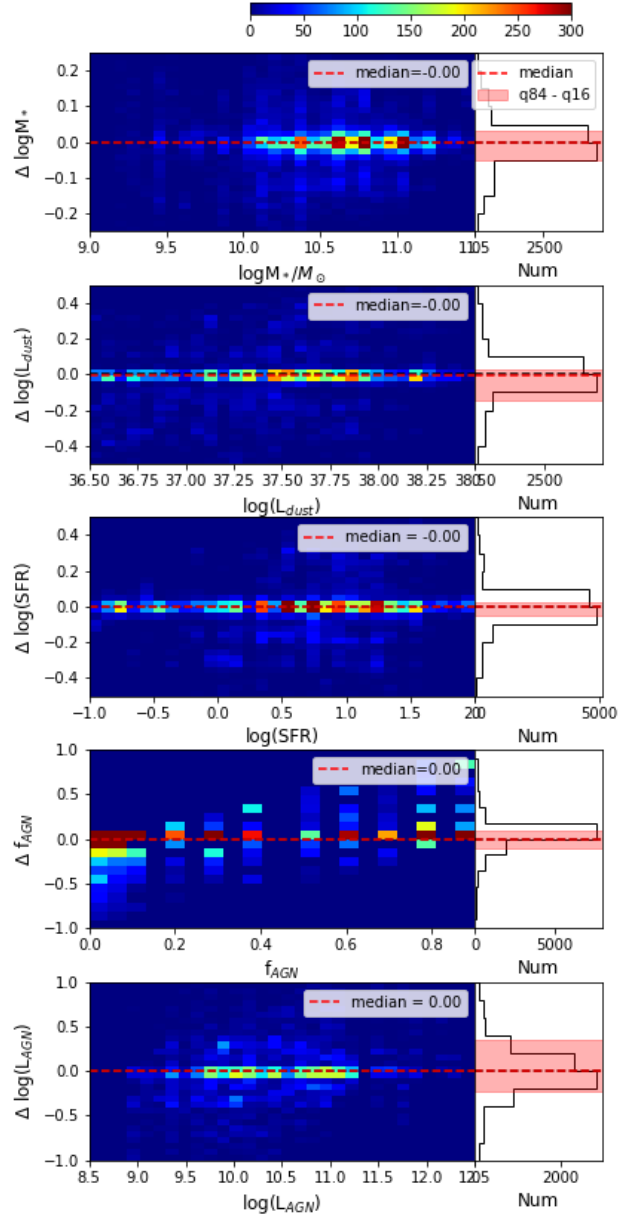}
    \caption{Comparison with physical quantities derived from 100 mocks of each radio-IR galaxy compared to the real catalog. \textit{Left}: 2D histograms of the difference of quantities estimated in mock versus that estimated in the real catalog, as a function of the real quantities, colour coded by the number of mocks in the 2D box. The same colour convention is used for all right panels as shown in the top of figure. From top to bottom panels are stellar mass, log($L_\text{dust}$), SFR, \fAGN~and \LAGN. The median and 16th/84th percentiles of the differences are listed in the legend. \textit{Right}: Histogram of the difference in the left panel. The median of the difference is marked in the red horizontal dashed line with the shaded red region corresponding to the spread of 16th and 84th percentiles. The median values of differences are 0, suggesting small uncertainties of these quantity.}
    \label{fig:mock_cigale}
\end{figure}

In this section, we adopt the procedure of generating mock catalogues in CIGALE, in order to check the reliability of estimated parameters. 
In brief, the mock photometric data for each object is modified by adding a value taken from a Gaussian distribution using the photometric error as the standard deviation. Then a mock catalog is obtained by running CIGALE on the same parameter space and same method as the real catalog (see \citealp{Boquien2019} for more details). In addition, we generate 100 mocks for each radio-IR galaxy. 
Figure \ref{fig:mock_cigale} shows the differences ($\Delta$ = mock - real) in \lm, \lsfr, \LIR, \fAGN, \LAGN~derived from mock catalogues and those derived from the real catalog as a function of real values and their histograms in the right panels. The median values of $\Delta$\lm, $\Delta$\lsfr), $\Delta$\LIR, $\Delta$\fAGN~and $\Delta$\LAGN~are $\sim$ 0. 

The largest difference is shown in the \fAGN~panel. At low~\fAGN~end, the \fAGN~from mocks are lower than the real value. The differences appear to increase as increasing \fAGN, indicating that CIGALE may underestimate the low~\fAGN~and overestimate high \fAGN. The scatter is large with differences from -1 to +1; however, the median is 0 $\pm$ 0.1 with uncertainty on 16th/84th percentiles. Therefore, we trust the determination of \fAGN~to a difference of 0.2 based on the 16th/84th percentiles. We note that the separation of our AGN sub-samples is larger than this reliable difference and the median \fAGN~of our low- and high-\fAGN~sub-samples is 0.2 and 0.8, respectively. As such, this lack of precision to $\pm$0.2 in \fAGN~does not affect any result shown in the paper. In fact, we included the uncertainty of \fAGN~in Section \ref{sec:agn_lum}, which do not change our result. 

\begin{table}
	\caption{Depth of Imaging Data}
	\label{tab:image_depth}
	\begin{threeparttable}
	    \begin{tabular}{lccccc} 
		    \hline
		    \hline
	    	Field & MIPS\tnote{1} & PSW\tnote{2} & PMW & PLW & VLA \tnote{3} \\
	    	Wavelegth & 24$\mu$m & 250$\mu$m & 350$\mu$m & 500$\mu$m & 1.4GHz \\
	    	Units & ($\mu$Jy) & (mJy) & (mJy) & (mJy) & ($\mu$Jy) \\
		    \hline
		    RXJ1716 & 65.4 (33.6) & - & - & - & 15.2\\
		    RXJ1821 & 39.7 (11.4) & - & - & - & 9.3\\
		    SG0023 & 4.1 (0.8) & 12.4 & 18.9 & 22.4 & 13.9\\
		    SC1604 & 9.4 (3.6) & - & - & - & 9.30\\
		    XLSS005 & 184.6 (90.1) & 5.9 & 13.5 & 14.2 & 14.1 \\
		    SC0910 & 74.4 (29.9) & 8.5 & 11.6 & 14.6 & 14.7 \\
		    RXJ1053 & 41.2 (16.0) & 6.7 & 15.3 & 15.1 & 10.2\\
		    Cl1350 &  & - & - & - & 9.8 \\
		    Cl1429 &  & - & - & - & 17.1 \\
         	\hline
		    \hline
	    \end{tabular}
	    \begin{tablenotes}
	    \item[1] 3$\sigma$ (1$\sigma$) point source completeness limit. 
	    \item[2] 1$\sigma$ detection limit of 250, 350, 500$\mu$m Herschel/SPIRE images as described in Section \ref{sec:FRIobs}. Note that the reported detection limits represent the upper limits in these highly confused images. 
	    \item[3] RMS sensitivity for final image associated with all data for that pointing. Note that Radio sources are selected with a significance $\geq 4 \times$RMS in at least one of the radio integrated or peak flux. 
	    \end{tablenotes}
   \end{threeparttable}
\end{table}


\bsp	
\label{lastpage}
\end{document}